\documentclass{article}
\usepackage{PRIMEarxiv}
\usepackage[utf8]{inputenc} 
\usepackage[T1]{fontenc}    
\usepackage{hyperref}       
\usepackage{url}            
\usepackage{booktabs}       
\usepackage{amsfonts}       
\usepackage{nicefrac}       
\usepackage{microtype}      
\usepackage{lipsum}
\usepackage{fancyhdr}       
\usepackage{graphicx}       
\graphicspath{{media/}}     
\usepackage{amsmath}
\usepackage{algorithm}
\usepackage{subcaption}
\title{Scenic Routes With Weighted Points in 2D}

\author{
  Vijayraj Shanmugaraj \\
  Data Science and Analytics Center\\
  International Institute of Information Technology, Hyderabad\\
  \texttt{vijayraj.shanmugaraj@alumni.iiit.ac.in} \\
   \And
  Lini Thomas \\
  Data Science and Analytics Center\\
  International Institute of Information Technology, Hyderabad\\
  \texttt{lini.thomas@iiit.ac.in} \\
  \And
  Kamalakar Karlapalem \\
  Data Science and Analytics Center
  International Institute of Information Technology, Hyderabad\\
  \texttt{kamal@iiit.ac.in} \\
}

\begin{document}
\maketitle

\begin{abstract}
In a given 2D space, we can have points that have different levels of importance. One would prefer viewing those points from a closer/farther position in accordance with their level of importance. A point in 2D from where the user can view two given points in accordance with his/her preference of distance is termed as a scenic point.
We develop the concept of scenic paths in a 2D space with respect to two points which have weights associated with them. Subsequently, we propose algorithms to generate scenic routes a traveller can take, which cater to certain principles which define the scenic routes. Following are the contributions of this paper: (1) mathematical formulation of a scenic point, (2) introduction of scenic routes formed by such scenic points in two-class point configurations in 2D spaces, and (3) design of scenic route generation algorithms that fulfil certain defined requirements.
\end{abstract}


\section{Introduction}\label{sec1}

Humans possess an intrinsic sense of what scenic beauty is, and how to compose various elements they perceive within a frame. The question in hand at the moment is whether one would be able to objectively define optimal positions to capture scenic beauty with respect to points in 2D space. Our work intends to do the same by providing tangible definitions of what constitutes a scenic point, and devise routes to travel over paths that constitute these scenic points.

\subsection{Motivation}
In a given 2D space, we can have points that have different levels of importance. One would prefer viewing those points from a closer/farther position in accordance with their level of importance in order to obtain a view as scenic as possible.

For instance, the scenic beauty of a particular point in a forest may be determined by the spread of a tree present on that point. Another factor that may determine the scenic importance of a point could be the perceived heights of the major landmarks present at that point. Consider two major landmarks, one being comparatively shorter than the other. If we try focusing on the taller landmark alone, we may end up with the shorter landmark occupying a smaller fraction of the frame, and hence would have comparatively poorer resolution of details than the larger one. If we try focusing on the shorter landmark alone, we may effectively end up cropping a part of the taller monument, effectively spoiling the aesthetic of the photograph/video. One may want to keep the shorter landmark in the photographic/videographic frame with minimal loss in resolution of its details, while keeping the taller landmark evidently large enough in the scene. One solution to this issue is to frame the landmarks in a manner such that both of them appear to be the same height, which helps us strike a middle ground while accommodating the two landmarks into the frame.
\begin{figure}
\centering
\includegraphics[width=\columnwidth]{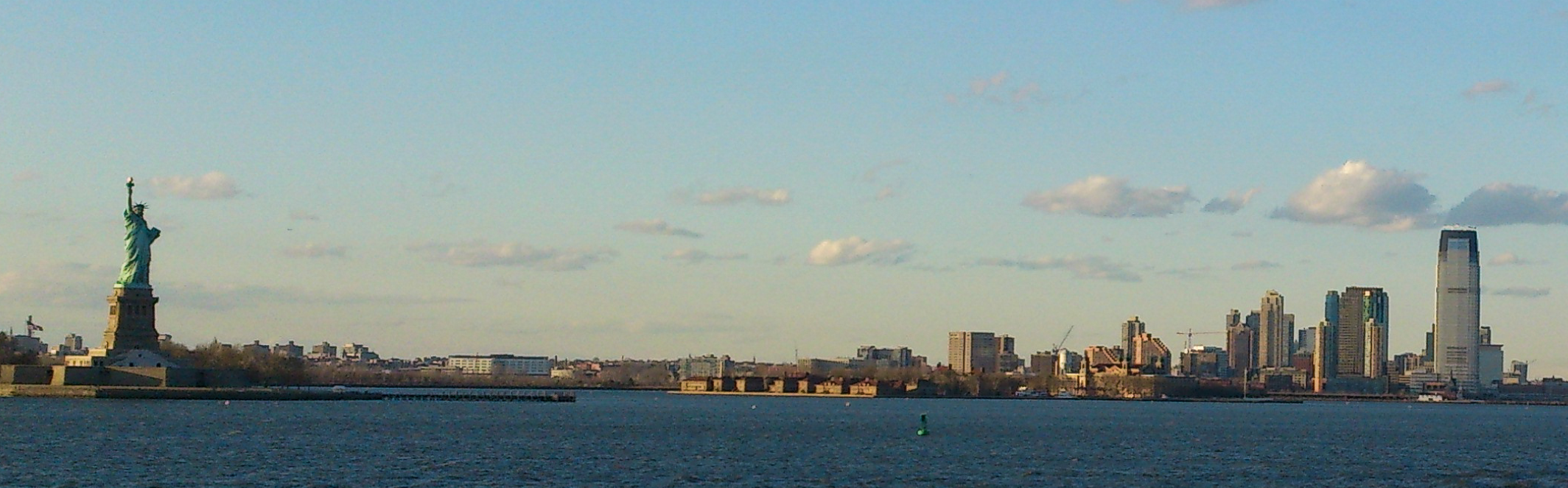}
\caption{A scenic image of the Statue of Liberty with the 30 Hudson Street building (the tallest tower on the right) among other skyscrapers\cite{pict2}, licensed under CC2.0}
\label{StatLib}
\end{figure}

For example consider Figure \ref{StatLib}. We can see that the Statue of Liberty monument and the 30 Hudson Street building in the background are nearly the same size. This photo composes both monuments in the frame without any one of them significantly overpowering the other (Note that the 30 Hudson Street building is around 238m tall, whereas the Statue of Liberty monument is 93m tall).

We develop the concept of scenic paths in a 2D space with respect to two points which have weights associated with them. These weights may signify the relative importance of a given point, or a characteristic of an object on said point that we want to emphasize upon, for example, the height or the total visible surface area of an object.

Subsequently, we propose algorithms to generate scenic routes a traveller can take that adhere to certain requirements that allow the traveller to cover all scenic views in an efficient manner.

\subsection{Paper Contribution and Organization}
Following are the contributions of this paper: (1) mathematical formulation of a scenic point, (2) introduction of scenic routes formed by such scenic points in two-class point configurations in 2D spaces, and (3) design of scenic route generation algorithms that fulfil certain defined requirements.

In Section \ref{BGSection}, we discuss the concept of apparent height and base the formulation for scenic routes on this concept. We go on in Section \ref{SRsection} to understand what a scenic route is, what its requirements are, and the motivations behind the same. In Section \ref{SRA}, we present three algorithms that generate scenic routes, and discuss the the results. In Section \ref{EVALSR} and \ref{Examples}, we finally present some synthetic and real world data sets on which the algorithms were tested on and present the results for the same.

\section{Background}
\label{BGSection}
\subsection{Apparent/Perceived height}
    \begin{figure}
    \centering
      \includegraphics[width=0.75\columnwidth]{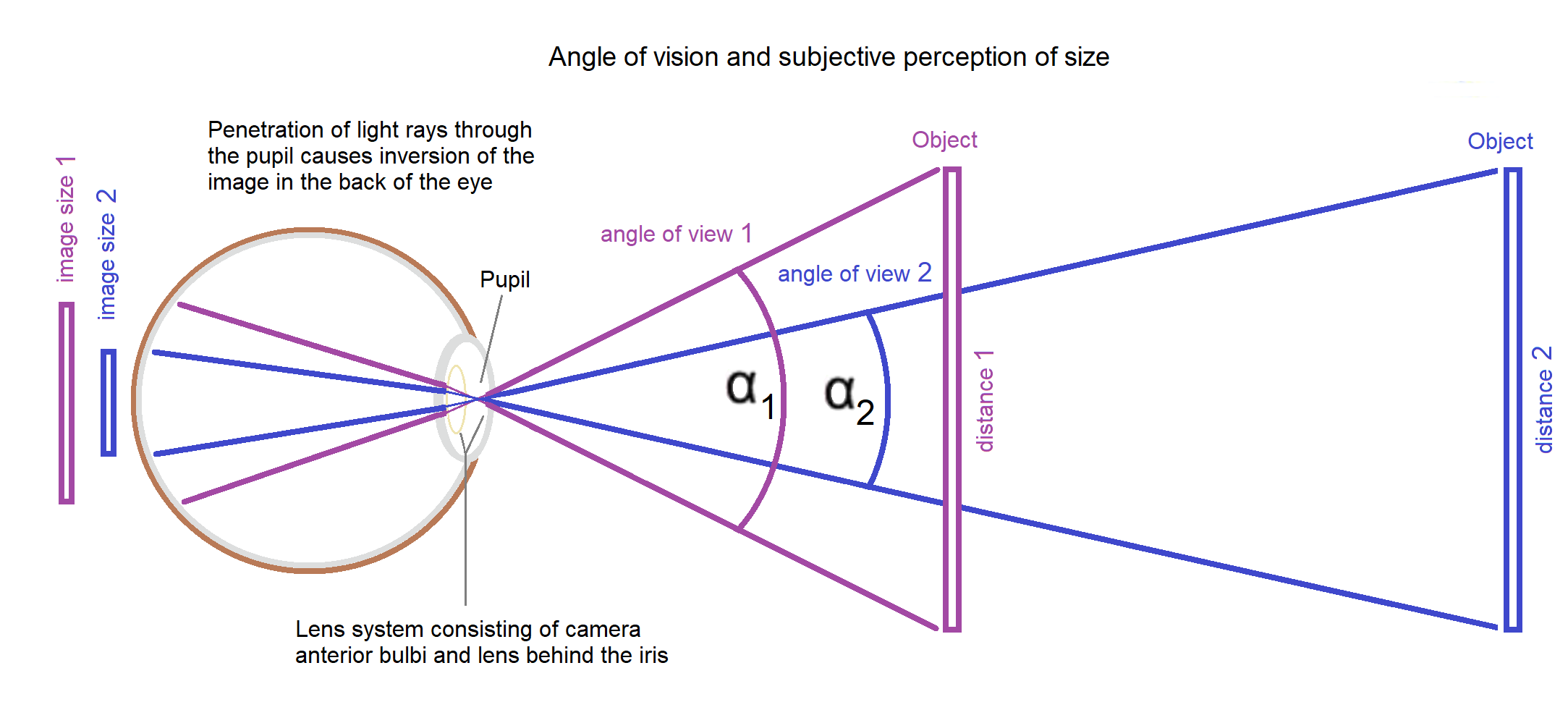}
    \caption{Visual angle of an object, and how distance of the object from the eye affects apparent size\cite{pict}, licensed under CC0}
    \label{fig-eye}
    \end{figure}

In this section, we discuss the concept of apparent height and formulate the condition for a point to be scenic with this as the backdrop.

The size of an object as perceived by the eye depends upon the size of the image it forms on the retina, which in turn depends on the \textbf{visual angle}. The visual angle is the angle a viewed object subtends at the eye as shown in Figure \ref{fig-eye}.

As evident from the Figure \ref{fig-eye}, an object closer to the eye subtends a larger visual angle at the pupil resulting in the object looking larger. Let us consider the case of a human looking at a tower and derive the expression for the tangent of the visual angle subtended by the tower on the person's eye.

\begin{figure}
\centering
\includegraphics[scale=0.2]{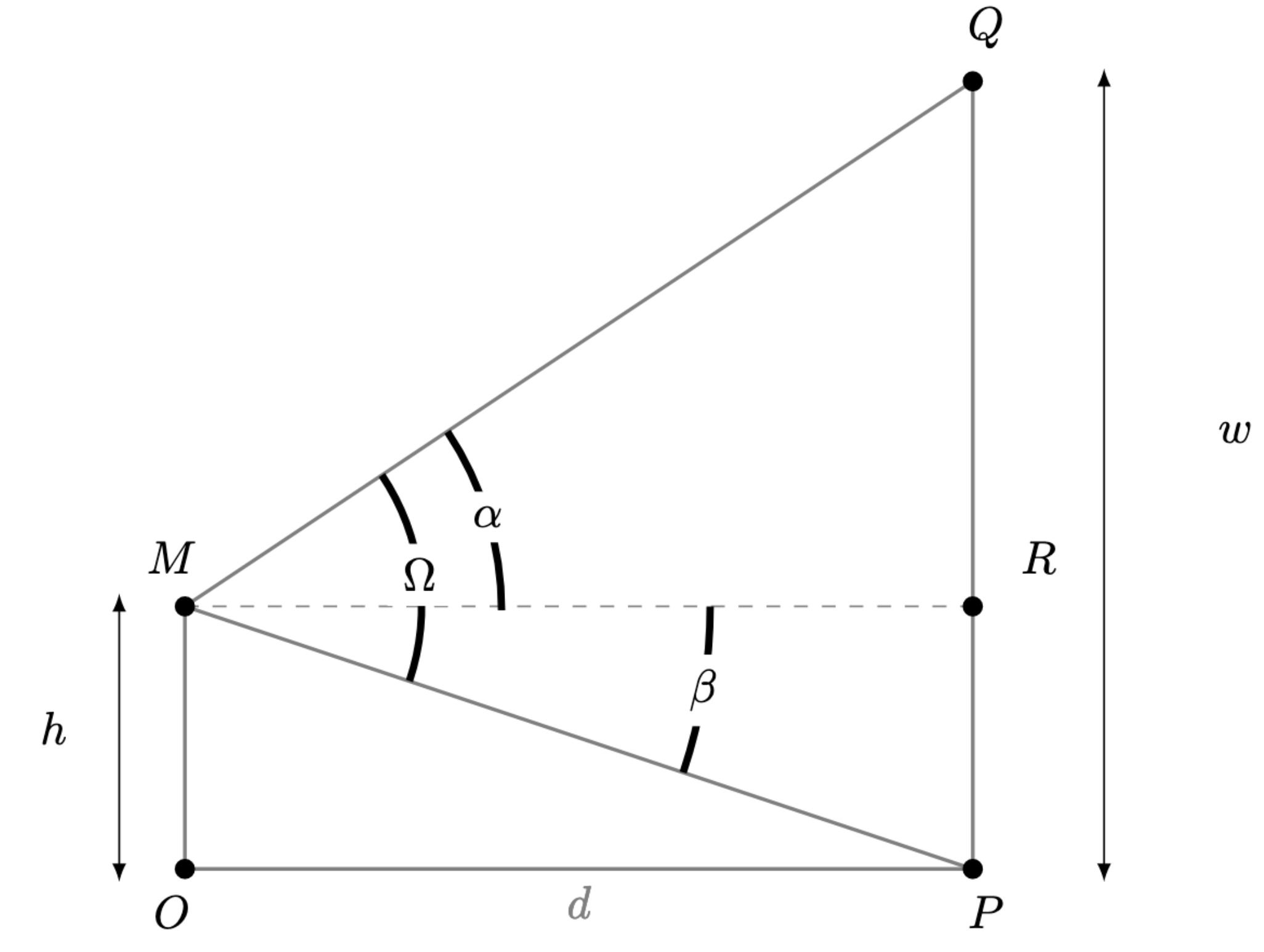}
\caption{A person (MO) viewing a tower (QP) from a distance of d units} 
\label{fig:mantower}
\end{figure}

Consider Figure \ref{fig:mantower}, where a person (represented by the line segment $MO$) is looking at a tower (represented by the line segment $QP$). Say that the person $MO$'s eye level is $h$ units from the ground and the tower $QP$'s height is $w$ units. The distance $OP$ between the person's feet and the tower's base is $d$. Then,

\begin{align*}
    MO = PR = h,\; OP &= d,\; PQ = w\\
    \angle QMP = \Omega,\; \angle QMR &= \alpha,\; \angle RMP = \beta
\end{align*}

The visual angle is $\Omega = \alpha + \beta$.

\begin{align*}
    \tan(\Omega) &= tan(\alpha + \beta)\\ &= \frac{\tan(\alpha)+\tan(\beta)}{1-\tan(\alpha).\tan(\beta)}\\
\end{align*}

Since $\tan(\alpha) = (w-h)/d$ and $\tan(\beta) = h/d$,

\begin{align*}
    \tan(\alpha + \beta)&= \frac{\frac{w}{d}}{1-\frac{(w-h)h}{d^2}}\\
    tan(\Omega) &= \frac{wd}{d^2-(w-h)h}\\ &= \frac{d}{\frac{d^2}{w}-(1-\frac{h}{w})\frac{h}{w}}
\end{align*}
If we assume that $w >> h$ i.e the landmarks are much taller than the human, or the human tries to view the tower by keeping his eyes at ground level, $h/w \rightarrow 0$. Hence,
\begin{align*}
    tan(\Omega) &\approx \frac{w}{d}
\end{align*}
Say the visual angle of a tower $t_1$ with height $w_1$ at a distance $d_1$ from the person is $\Omega_1$, and the visual angle of a tower $t_2$ with height $w_2$ at a distance $d_2$ from the person is $\Omega_2$. Now for their apparent heights to be equal, we require $\Omega_1 = \Omega_2$. Hence,

\begin{align*}
    tan(\Omega_1) &= tan(\Omega_2)\\
    \frac{w_1}{d_1}&=\frac{w_2}{d_2}\\
    w_1d_2 &= w_2d_1
\end{align*}

Any point that satisfies the above relation for the two landmarks is a scenic point. A collection of continuous scenic points that satisfies the above condition for a particular pair of landmarks constitute a scenic path.

Although this formulation of the condition for a scenic point is based on the concept of apparent heights, we can use the same formulation for other usecases by assigning weights to points accordingly.

\subsection{Scenic points and paths}
\begin{figure}[h]
\begin{center}
\includegraphics[width=0.5\columnwidth]{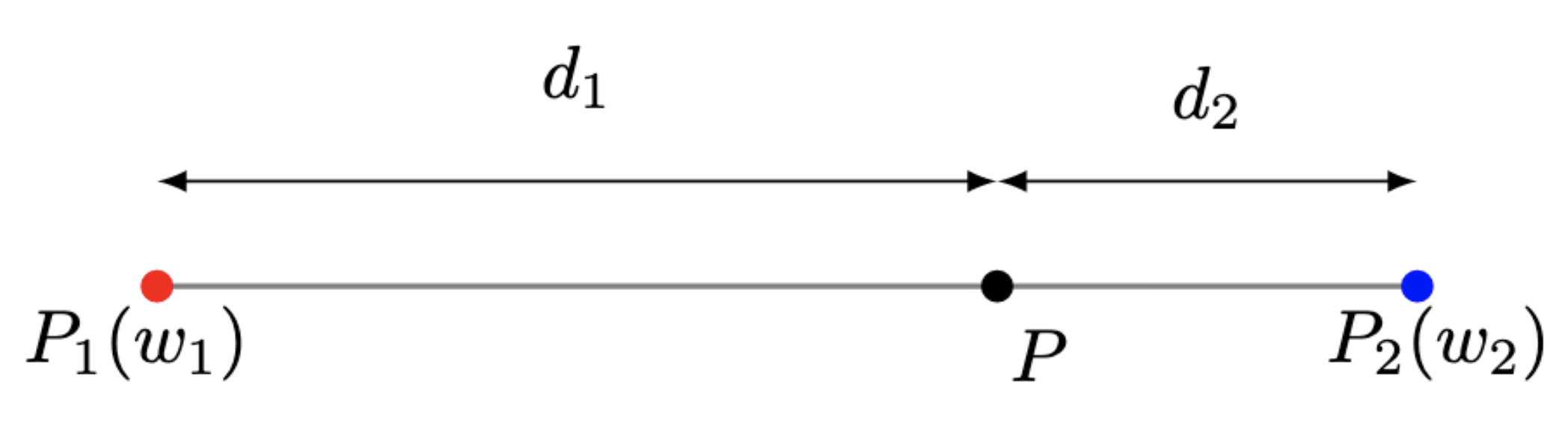}
\end{center}
\caption{A scenic point (in this case $w_2 < w_1$)}
\label{fig:locus}
\end{figure}

Consider a rectangular area with n red and m blue points, each point having its own weight. The apparent weight of a given weighted point $P_1$ with respect to another point $P_2$ is the ratio of the weight of the point $P_1$ (say $W_{P1}$) to the distance between the two points i.e. the apparent weight is equal to $W_{P1}/|\overline{\rm P_1P_2}|$. A \textbf{scenic point} is a point from where the apparent weights of the particular red point and a particular blue point are equal. Hence, the scenic point corresponding to a particular pair of red-blue points satisfies the following condition for a red point and a particular blue point (see Figure \ref{fig:locus}): 
\begin{align*}
    w_1d_2 = w_2d_1
\end{align*}

where $w_1$ and $w_2$ are the weights of the red point and the blue point respectively ($w_1 > 0$, $w_2 > 0$, and $d_1$ and $d_2$ are the distances of the scenic point from the red point and the blue point respectively).
A path on which each point is scenic is termed as a scenic path i.e. the locus of the point P in the above diagram would result in a scenic path.

Consider Figure \ref{fig:locus}. Let there be a landmark at $P_1(x_1, y_1)$ with weight $w_1$ and another landmark at $P_2(x_2, y_2)$ with weight $w_2$. Consider a point $P(x, y)$ that satisfies the condition $w_1d_2 = w_2d_1$, where $d_1$ is the distance of line segment $PP_1$ and $d_2$ is the distance of line segment $PP_2$. The locus of P will give us our desired route. Now,
\begin{align*}
    w_1d_2 &= w_2d_1\\
    \text{Let's say } \delta &= \frac{w_1}{w_2} \implies d_1 = \delta d_2\\
    \implies d_1^2 &= \delta^2 d^2_2
\end{align*}
We also have 
\begin{align*}
    \overline{PP_1} = d_1 &= \sqrt{(x - x_1)^2 + (y - y_1)^2}\\
    \overline{PP_2} = d_2 &= \sqrt{(x - x_2)^2 + (y - y_2)^2}
\end{align*}

Substituting for $d_1$ and $d_2$, we have
\begin{align*}
    (x - x_1)^2 + (y - y_1)^2 &= \delta^2 \big((x - x_2)^2 + (y - y_2)^2 \big)
\end{align*}
Simplifying, we have 
\vskip 0.2in
\begin{center}
\boxed{
    \!\begin{aligned}
    (\delta^2 - 1)x^2 -2(\delta^2 x_2 - x_1)x + (\delta^2 x^2_2 - x^2_1)\\
    (\delta^2 - 1)y^2 -2(\delta^2 y_2 - y_1)y + (\delta^2 y^2_2 - y^2_1) = 0
     \end{aligned}
    }
\end{center}
\vskip 0.2in
The locus of the above points is a circle, since a conic of the form $ax^2+2hxy+by^2+2gx+2fy+c=0$ is a circle if $a = b$, $a \neq 0$ and $h = 0$. These two criteria are satisfied if $\delta \neq 1$. In the case that $\delta = 1$, we have
\begin{align*}
    -2(x_2 - x_1)x + (x^2_2 - x^2_1)& \\ -2(y_2 - y_1)y + (y^2_2 - y^2_1)& = 0
\end{align*}
Simplifying,
\begin{align*}
    \bigg(y - \frac{y_1 + y_2}{2}\bigg) &= -\bigg(\frac{x_2 - x_1}{y_2 - y_1}\bigg)\bigg(x - \frac{x_1 + x_2}{2}\bigg)
\end{align*}
i.e. the curve will become the perpendicular bisector of the line segment joining the points $P_1$ and $P_2$.
\par
If $\delta \neq 1$, we can say that the scenic path produced will be a circle. For a generic circle with the formula $x^2+y^2-2gx-2fy+c=0$, the centre of the circle lies at $(g, f)$ and the radius of the circle is $\sqrt{g^2 + f^2 - c}$. It is possible that a circle may have an imaginary radius because of the negative term inside the square root, and hence, the circle would not exist in the real space. Because of this, we will have to prove that for any two points and any two weights, a circle will always exist.\par
In our original problem formulation, we have many free variables ($x_1, x_2, y_1, y_2, w_1, w_2$). Hence, we shall make the problem simpler without any loss of generality. We shall take a point $P_1$ at the origin $O$, and a point $P_2$ at the point $(d, 0)$ (i.e. on the x-axis at a distance $d$ from $P_1$, and let us consider the ratio of the weights to be $w_1/w_2 = \delta, w_1 >0, w_2 > 0, \delta \neq 1$. Any configuration of a pair of points can be converted into the above setup by appropriate geometric transformations (translation and rotations).

Now, substituting for $P_1$ and $P_2$ ($x_1 = 0, x_2 = d, y_1 = 0, y_2 = 0$) in the equation, we have
\begin{align*}
(\delta^2 - 1)x^2 + (\delta^2 - 1)y^2 -2\delta^2dx + \delta^2 d^2 = 0\\
x^2 + y^2 - \frac{2\delta^2d}{\delta^2 - 1}x + \frac{\delta^2 d^2}{\delta^2 - 1} = 0
\end{align*}
Comparing this equation with $x^2+y^2-2gx-2fy+c=0$, we have
\begin{align*}
    g &= \frac{\delta^2d}{\delta^2 - 1}\\
    f &= 0\\
    c &= \frac{\delta^2 d^2}{\delta^2 - 1}
\end{align*}
Now the radius of the circle would be
\begin{align*}
    r &= \sqrt{g^2 + f^2 - c}\\
    &= \sqrt{g^2 - c}\\
    &= \sqrt{\frac{\delta^4d^2}{(\delta^2 - 1)^2} - \frac{\delta^2 d^2}{\delta^2 - 1}}\\
    &= \sqrt{\frac{\delta^4d^2 - \delta^2 d^2(\delta^2 - 1)}{(\delta^2 - 1)^2}}\\
    &= \frac{\delta d}{|\delta^2 - 1|}
\end{align*}

Since, in this case, $\delta \neq 1$, and as $w_1 > 0$ and $w_2 > 0$, $\delta > 0$, and $d$, the distance between the points is a positive value, the above expression for $r$ is a real, positive value. Hence, we can see that for any configuration of a pair of points with any weights, we will always have a circle (if weights are unequal), or a line (if the weights are equal).
\section{Scenic Routes}
\label{SRsection}
\subsection{Formulation}
\label{formulationSection}
We have a set of red points $R$, and a set of blue points $B$. Each point $p$ can be defined by its coordinates $(x, y)$ and the weight $w_p$ of the point. Let us say that the paths generated by using the formula $w_{r_i}d_{b_j} = w_{b_j}d_{r_i}$ for all combinations of ($r_i$, $b_j$), $r_i \in R$, $b_j \in B$, is $C$. A scenic path is either a circle (for the curve generated from applying the scenic path condition on two points with unequal weights) or a line (on two points which have the same weight).

\begin{figure}
\centering
\includegraphics[scale=0.2]{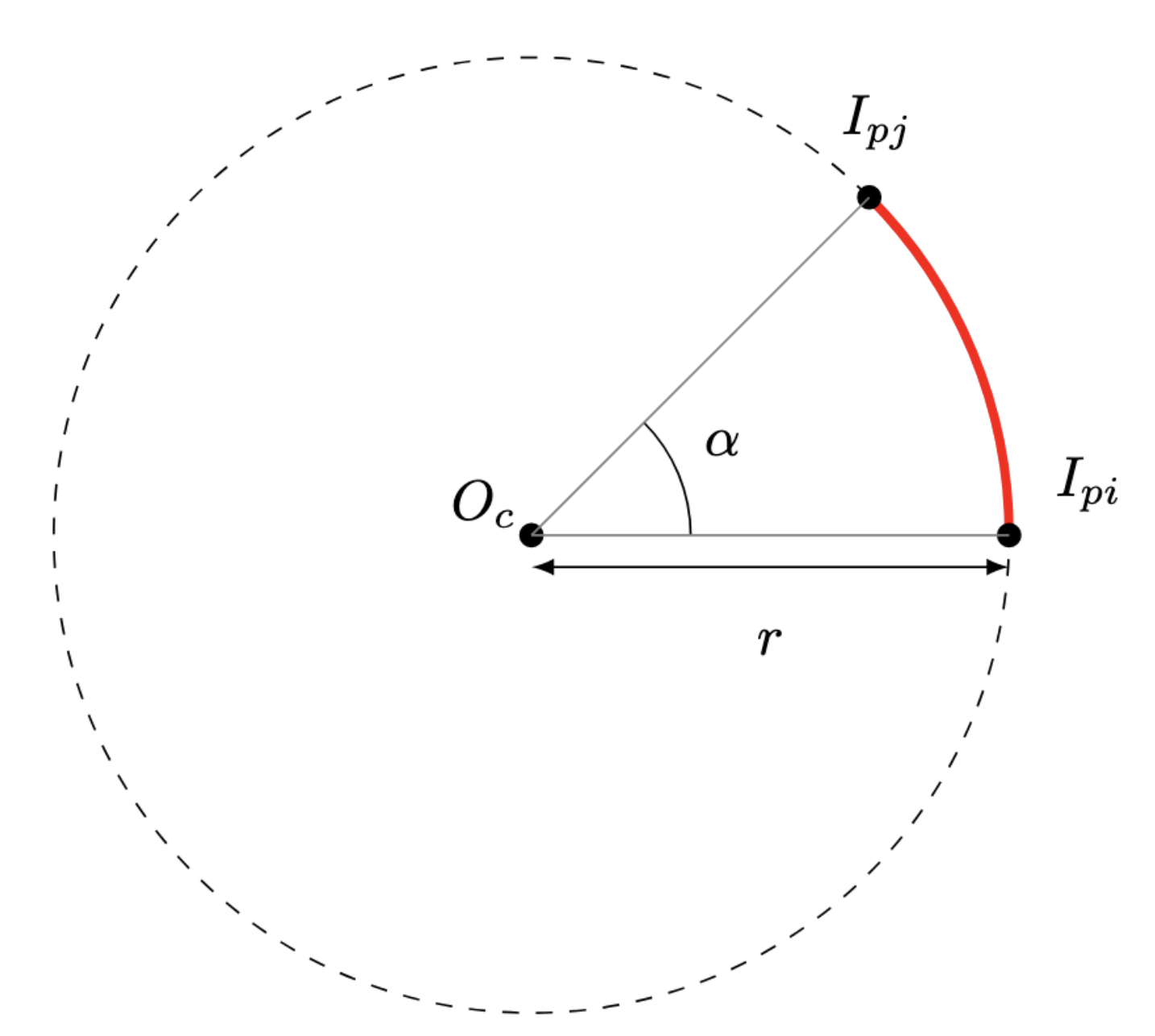}
\caption{An arc of a circle $c$ (with centre $O_c$) with its endpoints at $I_{Pi}$ and $I_{Pj}$}
 \label{fig:arcl}
 \end{figure}
 
A scenic path is partitioned into scenic edges through its intersection with other scenic paths. This set of partitioned edges that belong to a particular path $c \in C$ is $E_c$. The weight of each edge is the length of the edge. The length of an arc between two consecutive points $I_{Pi}$ and $I_{Pj}$ (see Figure \ref{fig:arcl}) on the same circle $c$ is $r.{\alpha}$, where r is the radius of the circle $c$ and $\alpha$ is the angle $\angle I_{Pi} O_c I_{Pj}$, where $O_c$ is the centre of the circle $c$.

Let us say that the set intersection points formed as a result of intersection of the curves is $I_P$. The edges between the intersection points of all such paths in $C$ form the edge set $E_P$ i.e. 

\begin{align*}
    \bigcup_{c \in C} E_c &= E_P
\end{align*}
\begin{figure}
    \centering
    \includegraphics[width=0.5\columnwidth]{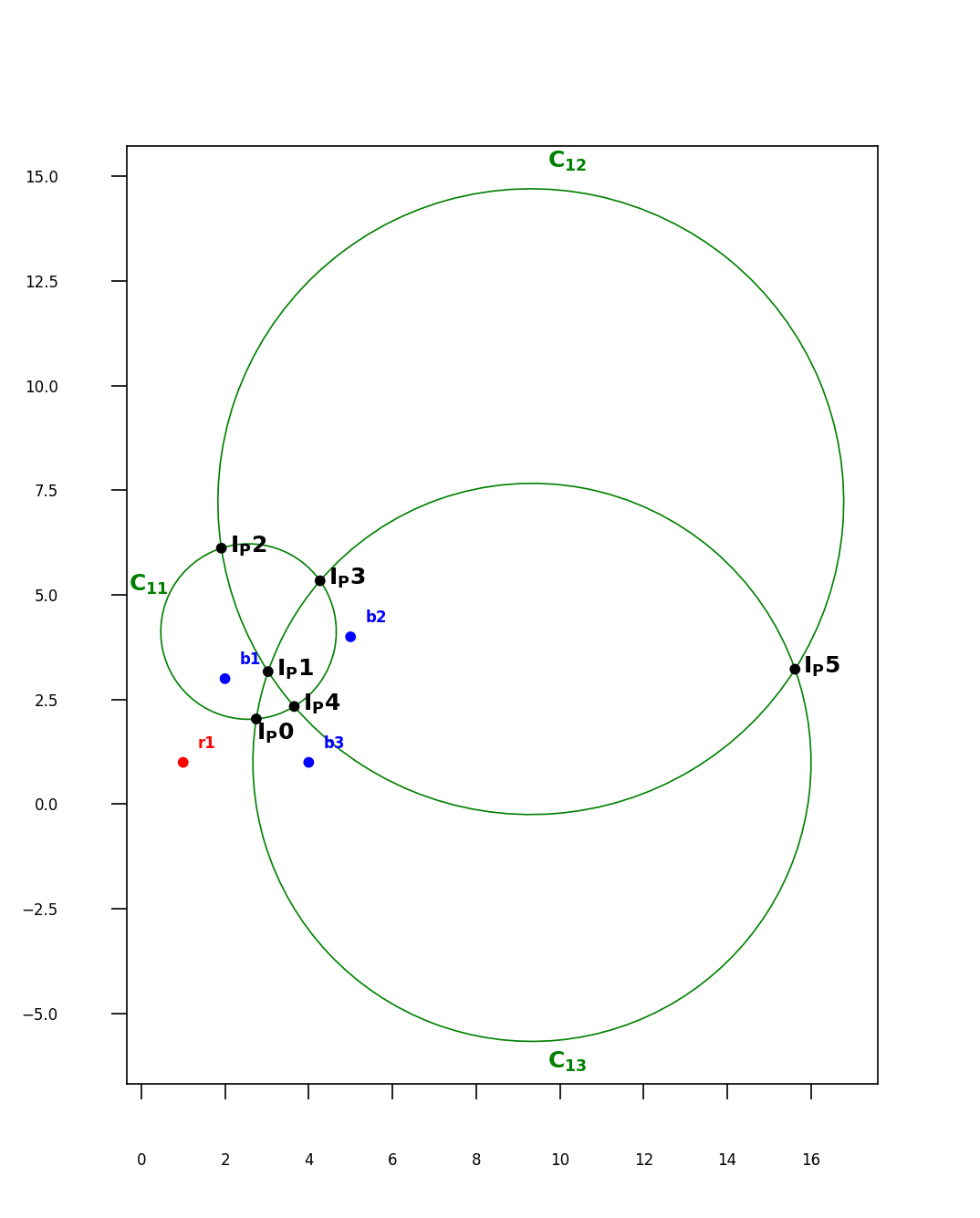}
    \caption{A sample graph with one red point, three blue points, the corresponding intersection points and the scenic paths (The points $r_1$, $b_1$, $b_2$ and $b_3$ have weights 2.5, 1.5, 1.8 and 2 respectively)}
    \label{fig:demo}
\end{figure}

The graph over all such intersection points $I_P$ is $G(I_P,E_P)$. For example, in Figure \ref{fig:demo}, we have $G(I_P,E_P)$ defined as follows: 

\begin{align*}
    \!\begin{aligned}
    R &= \{r_1\}\\
    B &= \{b_1, b_2, b_3\}\\
    C &= \{C_{11}, C_{12}, C_{13}\}\\
    I_P &= \{I_{P0}, I_{P1}, I_{P2}, I_{P3}, I_{P4}, I_{P5}\}\\
    E_P &= \{\overset{\Huge\frown}{I_{P0}I_{P1}}, \overset{\Huge\frown}{I_{P1}I_{P3}}, \overset{\Huge\frown}{I_{P3}I_{P5}}, \overset{\Huge\frown}{I_{P5}I_{P0}}, \overset{\Huge\frown}{I_{P0}I_{P2}}, \overset{\Huge\frown}{I_{P2}I_{P3}}, \overset{\Huge\frown}{I_{P3}I_{P4}}, \overset{\Huge\frown}{I_{P4}I_{P0}},     \overset{\Huge\frown}{I_{P4}I_{P1}}, \overset{\Huge\frown}{I_{P1}I_{P2}}, \overset{\Huge\frown}{I_{P2}I_{P5}}, \overset{\Huge\frown}{I_{P5}I_{P4}}\}
    \end{aligned}
\end{align*}

 Note that in some cases, we may have multiple sets of disjoint scenic paths i.e. we have a set of such sets of curves 
 \begin{align*}
 \sigma &= \{\sigma_i  : \sigma_i \subset C,\bigcup_{c_q \in \sigma_i, q \neq p} \text{ intersection}(c_p, c_q) \neq \phi \text{ for any } c_p \in \sigma_i\}
 \end{align*} where intersection$(c_p, c_q)$ is the function that gives the intersection points for two given curves $c_p$ and $c_q$. The above mathematical definition essentially states that any given circle $c_p \in \sigma_i$ will always intersect with at least one other circle $c_q \in \sigma_i$ (Note that $\bigcup_i \sigma_i = C$, and that if we have two circles $c_{pi}$, $c_{qj}$ such that $c_{pi}\in \sigma_i$, $c_{qj}\in \sigma_j$, $i \neq j$, intersection$(c_{pi}, c_{qj}) = \phi$ i.e. circles belonging to two separate disjoint components do not intersect).
   
 \begin{figure}
 \begin{center}
    \includegraphics[width=\columnwidth]{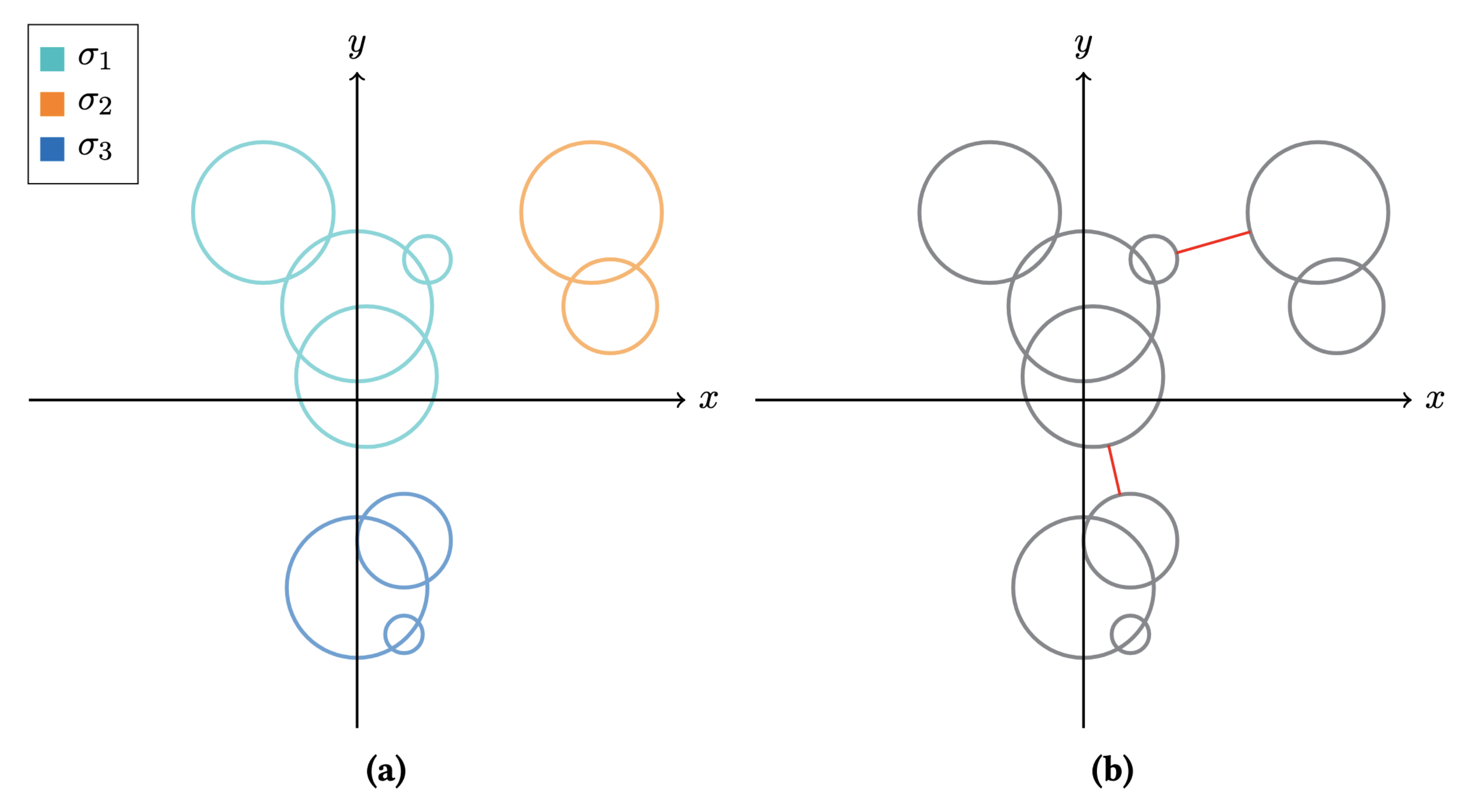}
     \end{center}
    \caption{(a) An example of a graph, whose $\sigma$ consists of three connected components $\sigma_1$, $\sigma_2$, $\sigma_3$, and (b) the edges (in red) added by the SCBA (See Section \ref{formulationSection}) in order to convert the above set of disconnected components into a connected graph}
    \label{sigma-graph}
\end{figure}
    
 Let us say that the set of all such disconnected components is $\sigma$. In such a case we connect individual disconnected components using the following procedure, which we shall name as the \textbf{Scenic Component Bridging Algorithm (SCBA)}. An example of a set of curves such that $|\sigma| = 3$, and the resulting edges added after running the SCBA is shown in Figure \ref{sigma-graph}.

 \par
    \begin{algorithm}
    \begin{flushleft}
     We first define a set $\Lambda$, which will hold a set of line segments which would be used to converted the disconnected graph into a connected one. For every pair ($\sigma_i$, $\sigma_j$) such that $\sigma_i \in \sigma$, $\sigma_j \in \sigma$, we do the following
     \end{flushleft}
    \begin{enumerate}
    \item 
        \begin{enumerate}
        \item Let us define $\epsilon_{ij}$ as the edge that belongs to $\sigma_i$ which is closest to $\sigma_j$. Take $\epsilon_{ij}$ and $\epsilon_{ji}$ and find the shortest straight line (say this is $l_{ij}$) between them.
        \item For every $l_{ij}$ we find the shortest line segment that is a part of $l_{ij}$ that has one end on any scenic edge in $\sigma_i$, and the other end on any scenic edge in $\sigma_j$. Say this line segment is $l$. We add $l$ to $\Lambda$.
    \end{enumerate}
    \textbf{Repeat this on $\sigma$ until $|\sigma| = 1$:}\par
    \item For all $l_k$ such that  $l_k \in \sigma$, we choose the line segment with the shortest length $l_{min}$. We then add this edge to $E_P$. If $l_{min}$ joins $\sigma_a$ and $\sigma_b$, we merge them into a single set with the addition of the edge $l_{min}$.
    \end{enumerate}
    \caption{Scenic Component Bridging Algorithm}
    \end{algorithm}
     
    The algorithms in this paper take $G(I_P,E_P)$ as input, and output a \textbf{scenic route}, which is a subgraph of $G(I_P,E_P)$. Let the set of nodes in such a route be $S \subseteq I_P$, and the edges within the route be $E \subseteq E_P$.
    
    We represent a scenic route as a graph using the notation $G(S,E)$. (Note that not all edges in a scenic route will be scenic, we may have some non-scenic edges that were added as a consequence of the Scenic Component Bridging Algorithm). 
    
\subsection{Properties of Scenic Routes}
\begin{figure}
    \centering
    \includegraphics[width=0.8\columnwidth]{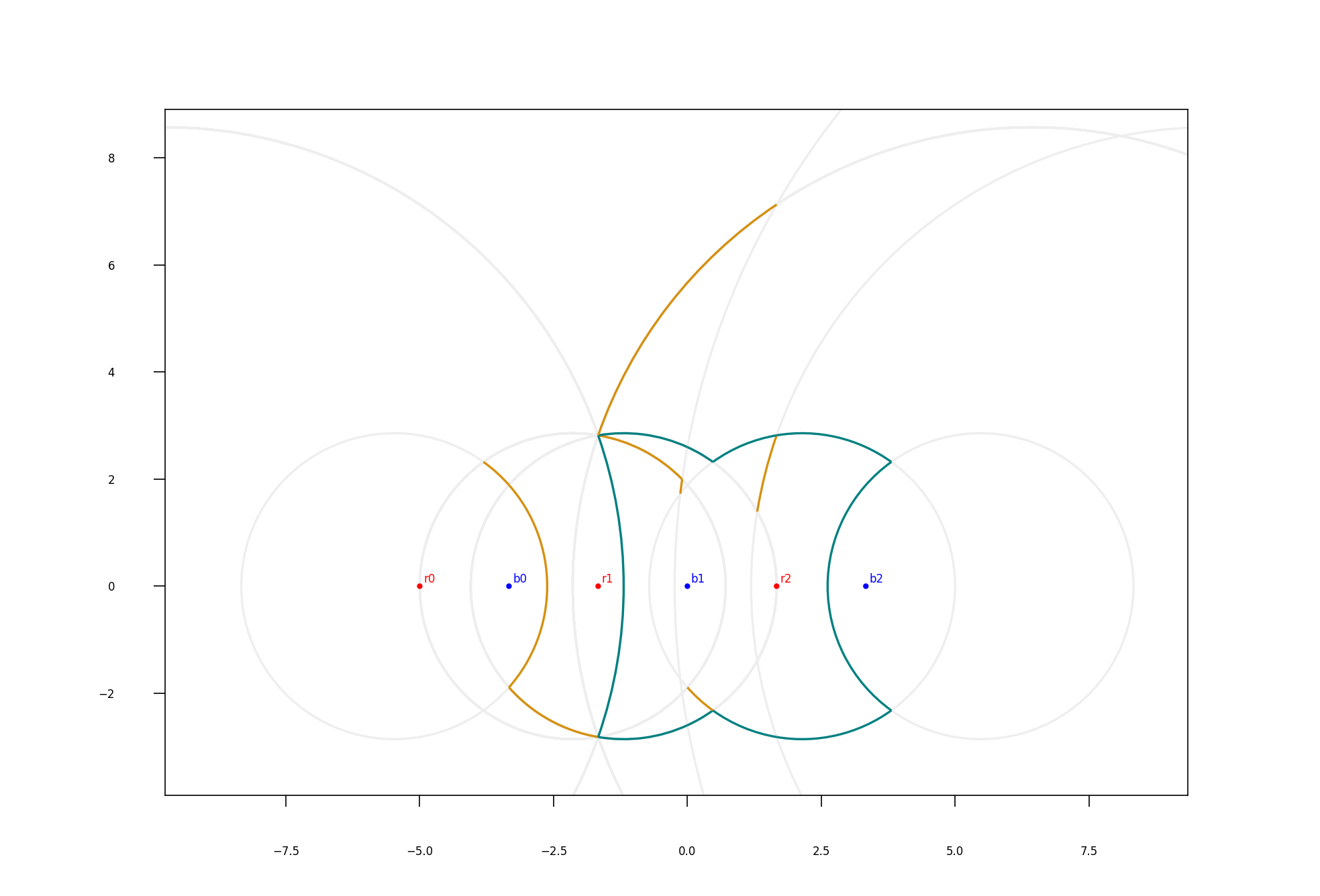}
    \caption{A scenic route with repeated edges highlighted in yellow}
    \label{fig:RE}
\end{figure}

Scenic routes have the following requirements, and the algorithms devised in this paper prioritize the requirements in the following order:
\begin{enumerate}
\item \textbf{Completeness:} Travelling on the route provided by the algorithm must allow one to have a view of a large number of red-blue pairs. It is desirable that a route would give a view of all red-blue pairs (i.e., all $|R|.|B|$ pairs). The ability to view a larger number of red-blue pairs on a scenic route would add to the scenic beauty offered by the route.
\item \textbf{Only scenic:} The route must consist of only scenic edges.
\item \textbf{Minimal route length:}  The total route length i.e the sum of all edges that are part of the route should be as low as possible
\item \textbf{Minimal repeated edges:} A route must contain a minimal number of repeated edges. Repeated edges are defined as stretches of paths that must be traveled more than once (repeated) to complete the entire route. These are paths whose one endpoint is free (not connected to any other edge i.e. the free endpoint has a degree of 1). (See Figure \ref{fig:RE}). Repeated edges come into play in order to produce a closed path to return to the starting point.
\item \textbf{Minimal number of edges:} A route must not have a large number of edges. Traveling on a route must allow for long, uninterrupted stretches of scenic points. In other words, there should not be a large number of direction changes within a route.
\end{enumerate}
 To better understand why the above requirements and the ordering of these requirements generates a preferred scenic route, consider a person walking on a scenic route.
\begin{enumerate}
\item Given that the red-blue points are points of interest, a person would want to scenically view as many (preferably all) red-blue points as possible, since they are all points of interest. (Req. 1)
\item As far as possible, the traveller would like to travel on a path that has a scenic view available. In the case of isolated scenic path components, the traveller would have to walk on a non-scenic route in order to reach a particular isolated component from the component they are on currently. (Req. 2)
\item The person would prefer travelling lesser distance, while covering all possible scenic paths. (Req. 3)
\item The traveller would not want to traverse on the same path repeatedly because repeated edges within scenic routes offer the same view multiple times, which unnecessarily increases the total length of the scenic route without offering any different novel scenic views. Hence they should be minimized. (Req. 4)
\item A person walking on a route would dislike a large number of direction changes, and would prefer longer edges on a particular path to not distract from the scenic beauty available. However, this requirement is last in the order of priority, since Req. 1 and Req. 3 would likely conflict with this requirement, since covering all scenic paths in the shortest route possible would require us to consistently switch paths, violating this requirement. (Req. 5)
\end{enumerate}

A route that fulfills the above requirements is a \textbf{preferred} scenic route. A route that fulfills fewer than five requirements is still a scenic route, but it would not be preferred.

While coming up with a scenic route, the importance of each of these requirements must be taken into consideration while making trade-offs in the case of conflicting requirements (for example requirements (1 and 3), and requirement 5).

\section{Scenic Route Algorithms}
\label{SRA}
As we mentioned in section \ref{formulationSection},
the problem is to generate a scenic route over the set of scenic paths and intersection points.
In this section, we design three 
scenic route generation algorithms and present their underlying motivations and objectives, and discuss our observations.
\par
Each algorithm takes the graph $G(I_P, E_P)$ as input, where $I_P$ is the set of all intersection points of the paths, and $E_P$ is the set of edges between these intersection points. The algorithms designed require knowing the shortest path that connects a pair of points. The shortest path is computed using the Floyd-Warshall all-pairs shortest path algorithm\cite{floyd1962algorithm}. APSP(a, b) denotes the distance between two points taken from the precomputed results of the Floyd-Warshall algorithm on $G(I_P, E_P)$. The following algorithms have been designed keeping completeness (Req. 1) as the primary objective in mind. 

If we have $n$ red points and $m$ blue points in a given 2D space, we will have $nm$ curves. The maximum number of intersection points we can have is $2.\binom{nm}{2} = nm(nm-1)$, since for each pair of curves (circles/lines), we can have a maximum of two intersection points. Without loss of generality, let us say that $n >= m$. In that case the maximum value of $nm(nm-1)$ that can be achieved is when $n = m$ i.e. the maximum number of curves that we can have is $n^2$, and hence, the maximum number of intersection points we can have for any $m$ such that $m <= n$ is $n^4 - n^2 \equiv O(n^4)$. 

In the subsequent sections, to calculate the complexity of each algorithm, we use this parameter $n$ (i.e. the maximum of the number of red points/the number of blue points, which essentially translates to the total number of landmark points, since $n + m <= max(2m, 2n) = 2n$ (in our case), and $O(2n) \equiv O(n)$) to represent complexity. (Note that the APSP calculation would be of the order (Number of points)$^3$, which would be $O(n^{12})$, but we do not take that into account since our algorithms consider the that the APSP on our graph is run offline, and the result is used as an input to our algorithms.)

\subsection{All Curve Umbrella (ACU) algorithm}
The ACU algorithm bases its route on the shortest edge from each scenic path (i.e. the smallest edge from each $E_c$, $c\in C$). Since the endpoints of each edge are common intersection points between two or more curves, we are trying to enable the traveller to easily switch between scenic views with minimal travelling distance. This algorithm explains how these edges are connected to each other to form a route. We next present the initialization and the algorithm details.

We next present the initialization and the algorithm details.
\subsubsection{Initialization}
 \label{ACInit}
This step involves selecting the shortest edge from each scenic curve. We insert all these edges into E, and insert all the edge endpoints into S i.e.
\begin{enumerate}
    \item E: set comprising of the shortest edge belonging to each curve present in the set C of graph G($I_P, E_P$)
    \item S: set of endpoints of all edges \; $e : e \in E$
\end{enumerate}

\subsubsection{Algorithm Details}
 After the above initialization, we calculate the centroid (g) of the points in S. 
 \begin{align*}
     g &= \Bigg(\frac{\sum_{i} x_i}{| S |} ,  \frac{\sum_{i} y_i}{| S |}\Bigg), i \in [1, | S |], i \in \mathbb{N}
 \end{align*}
 Where $s_i = (x_i, y_i), s\in S$. ($x_i$ and $y_i$ are the x and y coordinates of a point $s_i$ that belongs to the set S).
 
  We take the edges and sort them in the order of the angle made by the line joining the midpoint of the edge endpoints and $g (x_g, y_g)$, and the line parallel to the X-axis.

 We iterate through the list E in this order, and we connect each edge in the list with the next edge by finding the closest endpoint from the next edge using the paths specified by the APSP array. By doing this, we try to build a circular route around a central point that provides the user with a systematic well-defined order to cover all the edges in E.

\begin{algorithm}
Execute steps 1-3 twice, with the edges arranged in clockwise and anticlockwise order, and pick the route that has the minimum path length out of the two results:
\begin{enumerate}
    \item Calculate the centroid of the points $g (x_g, y_g)$ using the aforementioned formula. 
    \item Arrange the edges in E in a circular manner around $g$ (we sort the points in S (in either ascending or descending order, depending on whether we are trying to arrange the edges in a clockwise or anticlockwise manner) with the anticlockwise angle formed between the midpoint of the edge in E and $g$, and the line $y = y_g$.) Let us call this sequence $\Gamma$. 
    \item Loop over the above sequence of edges and join them as follows: (In the following example, we will use the notation $e_{11}$ and $e_{12}$ to denote the endpoints of the edge we are currently dealing with, and $e_{21}$ and $e_{22}$ to denote the endpoints of the next edge in the sequence $\Gamma$)
    \begin{enumerate}
        \item \textbf{If the edge is the first edge of $\Gamma$:} We check all four combinations i.e. $e_{11}$-$e_{21}$, $e_{11}$-$e_{22}$, $e_{12}$-$e_{21}$, $e_{12}$-$e_{22}$, and we add the edges corresponding to the APSP between the pair of points closest to each other. 
        \item \textbf{If the edge is neither the first or last edge of $\Gamma$:} Without loss of generality, let us assume that the endpoint of the current edge that is connected to the previous edge is $e_{11}$. We then add the edges corresponding to the APSP between the points of the closest pair in $e_{12}$-$e_{21}$ and $e_{12}$-$e_{22}$.
        \item \textbf{If the edge is the last edge of $\Gamma$:} Without loss of generality, let us assume that the endpoint of the current edge that is connected to the previous edge is $e_{11}$, and the endpoint of the first edge in $\Gamma$ that is already connected is $e_{21}$. We then add the edges corresponding to the APSP between the pair $e_{12}$-$e_{22}$. 
    \end{enumerate}
\end{enumerate}
\caption{ACU Algorithm}
\end{algorithm}

\subsubsection{Observations}
Listing out the requirements from a scenic route, 
\begin{enumerate}
\item \textbf{Completeness:} Travelling on the route provided by the algorithm must allow one to have a view of a large number of red-blue pairs. It is desirable that a route would give a view of all red-blue pairs (i.e., all $|R|.|B|$ pairs). The ability to view a larger number of red-blue pairs on a scenic route would add to the scenic beauty offered by the route. In this paper, we adopt a stricter version of the completeness criterion, where the traveller has to traverse on at least one edge of the path pertaining to a given red-blue pair to consider the pair covered, in order to thoroughly enjoy a given scenic path. However, the user can always choose to relax this by considering the condition for counting a path to be visiting a point on the path.

\item \textbf{Only scenic:} This criterion is satisfied only if all the curves are connected i.e. there are no isolated set of curves. If such disconnected components exist, this algorithm will pick all non-scenic routes that enable all scenic paths to stay connected.
\item \textbf{Minimal route length:} The length of the route is minimal when compared to the total length of all paths combined. 
\item \textbf{Minimal repeated edges:} As the points of the edges are connected in a circular manner based around the centroid, the final route generated is unlikely to have a repeated edge. Hence, this condition usually is satisfied by this algorithm. The only case when a repeated edge would be generated is in the following scenario: 
\par
Consider the following setting below (See Figure \ref{fig:repeatedEdge} for reference). Let us say that the edge $AB$ is connected to the previous edge in $\Gamma$ via the path $AP_1$. Now since $A$ is not a free endpoint anymore, we must connect end $B$ to the next edge. Say the closer endpoint in the next edge to $B$ is $P_2$. Now, in the case when APSP($B$, $P_2$) = $\{AB\} \cup \{e : e \in AP_2\}$ i.e. the shortest path from $B$ to $P_2$ is from $B$ to $A$ and utilizes the edges in the path from $A$ to $P_2$. In such a case, a repeated edge is unavoidable. 

\begin{figure}
    \centering
    \includegraphics[scale=0.5]{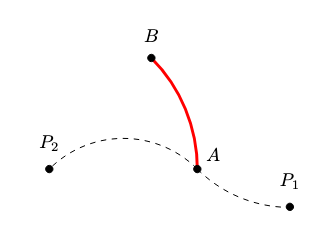}
    \caption{Diagram depicting the scenario for a repeated edge in the ACU algorithm}
    \label{fig:repeatedEdge}
\end{figure}

\item \textbf{Minimal number of edges:} The algorithm fails to deliver when it comes to Req. 5, since this algorithm tries to navigate between the shortest edges of each path. A short edge indicates that it is one of the edges with minimal distance which allows the traveller to switch from one scenic route to another. This in turn means that the traveller would have to make multiple switches, and go through multiple shorter edges in the interest of Req. 1 and 3; that is, the traveller will not get long, uninterrupted paths. 
\end{enumerate}

\subsubsection{Complexity}
The initialization of the All-curve algorithms requires the sorting of edges. A graph with $k$ nodes can have at most $k(k-1)/2 \equiv O(k^2)$ edges. And hence, the number of edges in our graph would be of the order of $O(n^8)$. Sorting $k$ edges would have a complexity of $k\log(k)$, and hence sorting the edges we have would have a complexity of $O(n^8\log(n^8)) \equiv O(n^8\log(n))$. We would be picking $O(n^2)$ edges from all these edges. Calculating the centroid of all the edge endpoints would be $O(n^2)$ i.e. linear order of the number of edges. Sorting them around the centroid would be an additional $O(n^2\log(n^2) \equiv O(n^2\log(n)$, and joining these edges would be of complexity $O(n^2)$ i.e. linear order of the number of edges (since joining any pair of edges would be O(1) as we have the APSP array with us). The final complexity of the algorithm is $O(n^8\log(n))$, since it tends to overpower the other steps' complexity.

\subsection{All Curve Convex Hull (ACCH) algorithm}
To address the previous algorithm's shortcomings, the All Curve Convex Hull algorithm works on the Convex Hull\cite{jarvis1977computing} of the set S rather than the endpoints of the edges in E. We connect the points of the hull with the APSP-specified path. While connecting the points with the edges in the APSP, we check whether the two endpoints of each edge have already been connected to each other by previously inserted edges. 
The initialization of this algorithm is the same as the one specified in section \ref{ACInit}.

\begin{algorithm}
\begin{enumerate}
 \item Generate the Convex Hull of S. Let the sequence of points given by the convex hull be $\Lambda$.
 \item Connect every two consecutive points in the sequence $\Lambda$ using the edges specified by the APSP between each of those two points. While joining the sequence $\Lambda$ using the paths specified by the APSP array, check if the two endpoints of each edge in the APSP are already connected, and add the edge only if they are not.
 \item If disconnected components remain in the new graph, connect each disconnected component with the closest disconnected component using the edges specified by the APSP.
\end{enumerate}
\caption{ACCH Algorithm}
\end{algorithm}

\subsubsection{Observations}
Listing out the requirements from a scenic route, 
\begin{enumerate}
\item \textbf{Completeness:} This algorithm is designed in such a manner such that it selects at least one edge from each scenic path. Hence, the completeness criterion is satisfied.
\item \textbf{Only scenic:} This criterion is satisfied only if all the curves are connected i.e. there are no isolated set of curves. If such disconnected components exist, this algorithm will pick all non-scenic paths (as specified by the Scenic Component Bridging Algorithm) that enable all scenic paths to stay connected.

\item \textbf{Minimal route length:} The length of the route is minimal when compared to the total length of all paths combined, and  the total route distance is of similar order to the of the length of the paths produced by the ACU algorithm.
\item \textbf{Minimal repeated edges:} This algorithm, however, has much more repeated edges than the ACU algorithm, due to the attempt to satisfy the minimal route length (Req. 3). We also connect disconnected edges in the end of step 2 in the algorithm with APSP. These disconnected edges along with the APSP path connecting the disconnected edges to the hull additionally contribute to the number of minimal edges in the graph.
\item \textbf{Minimal number of edges:} The algorithm has flaws similar to the ACU algorithm due to the same reasons, however performs better than ACU algorithm since it eliminates multiple edges that the ACU algorithm introduced. 

\end{enumerate}

\subsubsection{Complexity}
The initialization of the All-curve algorithms would have a complexity of $O(n^8\log(n))$, as seen in the previous algorithm. We would have $O(n^2)$ points in S (as we would be picking $O(n^2)$ edges). Calculating the convex hull of k points is $O(k\log(k)$, and hence the complexity of the convex hull of these $O(n^2)$ points would be would be $O(n^2\log(n^2)) \equiv O(n^2\log(n))$. In the worst case, we would have all the points in S in the convex hull, and hence joining the points of the hull would be of complexity $O(n^2)$ i.e. linear order of the number of points (since joining any pair of edges would be O(1) as we have the APSP array with us). The final complexity of the algorithm is $O(n^8\log(n))$, since it tends to overpower the other steps' complexity.

\subsection{Dense Point Expansion (DPE) algorithm}
In this algorithm, we aim to achieve completeness by choosing edges that are attached to well connected points i.e. points with a higher degree. A point with a high degree indicates that it would act as a hub between edges belonging to multiple scenic edges/circles, giving the traveller multiple path choices at every node. And hence, by choosing such points, we attempt to cover all possible scenic views with as minimal hops as possible, eliminating trivial edges in the route.

\par
\textbf{Initialization:} We take the set of all intersection points in the original graph $I_P$ and sort them in the descending order of their degree. If two points have the same degree, we sort them in the ascending order of the sum of edges that terminate at that point. We do this since sometimes, vertices of the same degree may be extremely far away from the main portion of the graph, occasionally leading to such vertices being picked, which in turn may result in the creation of longer paths. S and E are empty sets that will finally contain the final set of vertices and edges respectively.

\begin{algorithm}
\begin{enumerate}
    \item Pick the point with highest degree in $I_P$, say $I_0$, and for each edge terminating at $I_0$, check if the edge lies on a curve already covered by an edge in E. If not, add the edge to E, and add $I_0$ and the other endpoint of the edge in S.
    \item Remove $I_0$ from $I_P$, and repeat step 1 until there is at least one edge from each scenic pair in E
    \item Generate the Convex Hull of S. Let the sequence of points given by the convex hull be $\Lambda$. Connect every two consecutive points in the sequence $\Lambda$ using the edges specified by the APSP between each of those two points.
    \item If disconnected components remain in the new graph, connect each disconnected component with its closest disconnected component using APSP.
\end{enumerate}
\caption{DPE Algorithm}
\end{algorithm}
\subsubsection{Observations}
Listing out the requirements from a scenic route, 
\begin{enumerate}
\item \textbf{Completeness:} This algorithm is designed in such a manner such that it selects at least one edge from each scenic path. Hence, the completeness criterion is satisfied.
\item \textbf{Only scenic:} This criterion is satisfied only if all the curves are connected i.e. there are no isolated set of curves. If such disconnected components exist, this algorithm will pick all non-scenic routes that enable all scenic paths to stay connected.
\item \textbf{Minimal route length:} The DPE algorithm gives route lengths similar to the ACCH algorithm, with no clear winner. The similar path lengths might be due to the usage of the convex hull, since the DPE algorithm may end up picking some edges in common with the ACCH algorithm as a part of step 1, and plot a convex hull around these edges.
\item \textbf{Minimal repeated edges:} The DPE algorithm, however, seems to have a bit higher number of repeated edges than the former algorithms. This is probably because in this algorithm includes edges originating from points we iterate through before drawing out the convex hull, giving rise to a star-network-type subgraph, some of whose edges may have one free edge even after the components have been connected by the convex hull. Additionally, we connect disconnected edges in a similar fashion as ACCH algorithm. These two factors combined contribute to the higher edge percentage of the graph obtained. This also occasionally results in a large stretch of repeated edges as can be observed in some example datasets in the next section.
\item \textbf{Minimal number of edges:} The number of edges are similar to the number of edges in the graph produced by the ACCH algorithm. 
\end{enumerate}

\subsubsection{Complexity}
In the DPE algorithm, we sort the vertices in the graph, which would be of complexity  $O(n^2\log(n^2)) \equiv O(n^2\log(n))$. We would be picking $O(n^2)$ edges in the process of iterating through the vertices. In the worst case, we may have to iterate through all the vertices, and hence this operation would have a complexity of $O(n^2)$. Calculating the convex hull of k points is $O(k\log(k)$, and hence the complexity of the convex hull of these $O(n^2)$ points would be $O(n^2\log(n^2)) \equiv O(n^2\log(n))$. In the worst case, we would have all the points in S in the convex hull, and hence joining the points of the hull would be of complexity $O(n^2)$ i.e. linear order of the number of points (since joining any pair of edges would be O(1) as we have the APSP array with us). The final complexity of the algorithm is $O(n^2\log(n))$.

\section{Evaluation of Scenic Routes Algorithms}
\label{EVALSR}
We present the results of the algorithms on certain synthetic datasets, and the results of the algorithm run on 100 different randomly generated graphs:
(Note that the input given for the graph generation had four red points and four blue points, whose x and y coordinates lie in the range [-30, 30], and whose weights lie in the range [1, 50]. A graph (after disconnected components being connected by the SCBA algorithm) had 82.96 nodes, 176.96 edges and a total path length of 41302.61 on an average.)

For the synthetic datasets, all the red points have a uniform fixed weight $w_r$, and all the blue points have a uniform weight $w_b$. As seen in section 1.4, the only parameter that matters is the ratio of weights $w_b/w_r = \delta$. We will be labelling the graphs with the value of $\delta$ alone.

Note: In the graphs generated by the ACU and ACCH algorithms, we have marked some edges in magenta and some in bluish-green. The edges in magenta are the edges that were originally part of E in the all curve algorithm initialization in section \ref{ACInit}. The other edges that are in bluish-green are added as a part of subsequent steps in the graph building. The edges marked in yellow are repeated edges.

The results observed for the three algorithm are in line with the observations specified in the previous sections. While the ACU algorithm performs better at limiting the number of repeated edges, the ACCH and the DPE algorithms seem to do better at keeping the path length minimal.

\subsection{Experiment result data}
\begin{table}[h]
\begin{center}
\caption{List of abbreviations used in this section}\label{tab0}%
\begin{tabular}{cl}
\toprule
Abbreviation & Full form\\
\midrule
RL & Total Route Length\\
NoE & Number of Edges in route\\
NoRE & Number of Repeated Edges in route\\
RE\% & Repeated Edge Percentage\\
 \bottomrule
\end{tabular}
\end{center}
\end{table}

\begin{table}[H]
\begin{center}
\caption{Random graph experiment results (over 100 graphs, avg. number of nodes: 82.96, avg. number of edges: 176.96, avg. total edge length of graph: 41302.61)}\label{tab1}
\begin{tabular} {lllll}
\toprule
Algorithm & RL & NoE & NoRE & RE\%\\
\midrule
DPE & 194.71& 41.59& 11.12& 27.07\\
ACU & 218.29& 46.13& 3.77& 8.03\\
ACCH & 209.99& 43.33& 10.5& 24.17\\
 \bottomrule
 \end{tabular}
\end{center}
\end{table}

\begin{table}[H]
\begin{center}
\caption{The same Random graph experiment results in Table \ref{tab1}, presented with DPE as baseline ("R. X" stands for the ratio of the average value of the metric X of the routes produced by the algorithm in question to the average value of the metric X of the routes produced by the DPE algorithm))}\label{tab2}
\begin{tabular} {lllll}
\toprule
Algorithm & R. RL & R. NoE & R. NoRE & R. RE\%\\
\midrule
DPE & 100& 100& 100& 100\\
ACU & 114.92& 112& 40.74& 37\\
ACCH & 110.46& 104.73& 94.42& 89.28\\
 \bottomrule\end{tabular}
\end{center}
\end{table}

\begin{figure}
\begin{tabular}{cc}
\subcaptionbox{\label{g1}}{\includegraphics[width = 0.5\columnwidth]{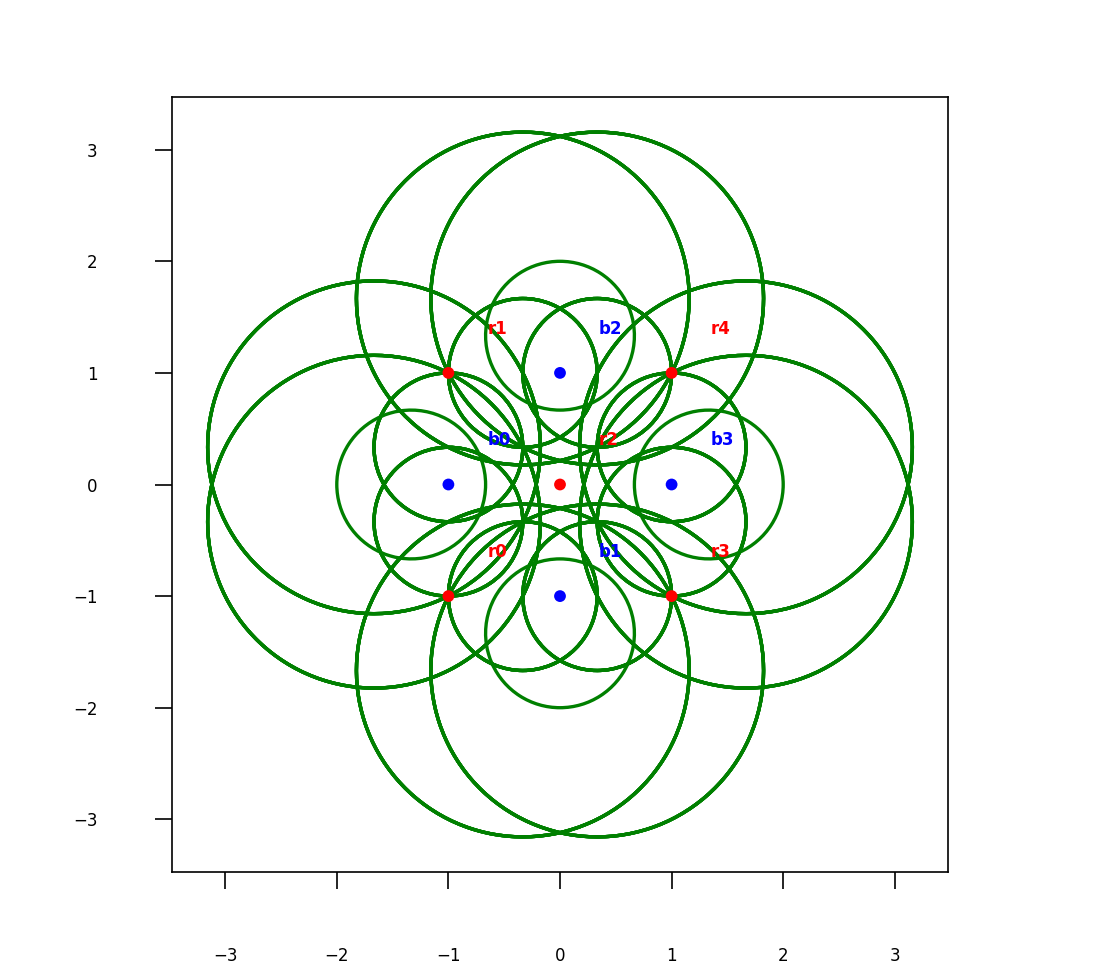}} &
\subcaptionbox{\label{g2}}{\includegraphics[width = 0.5\columnwidth]{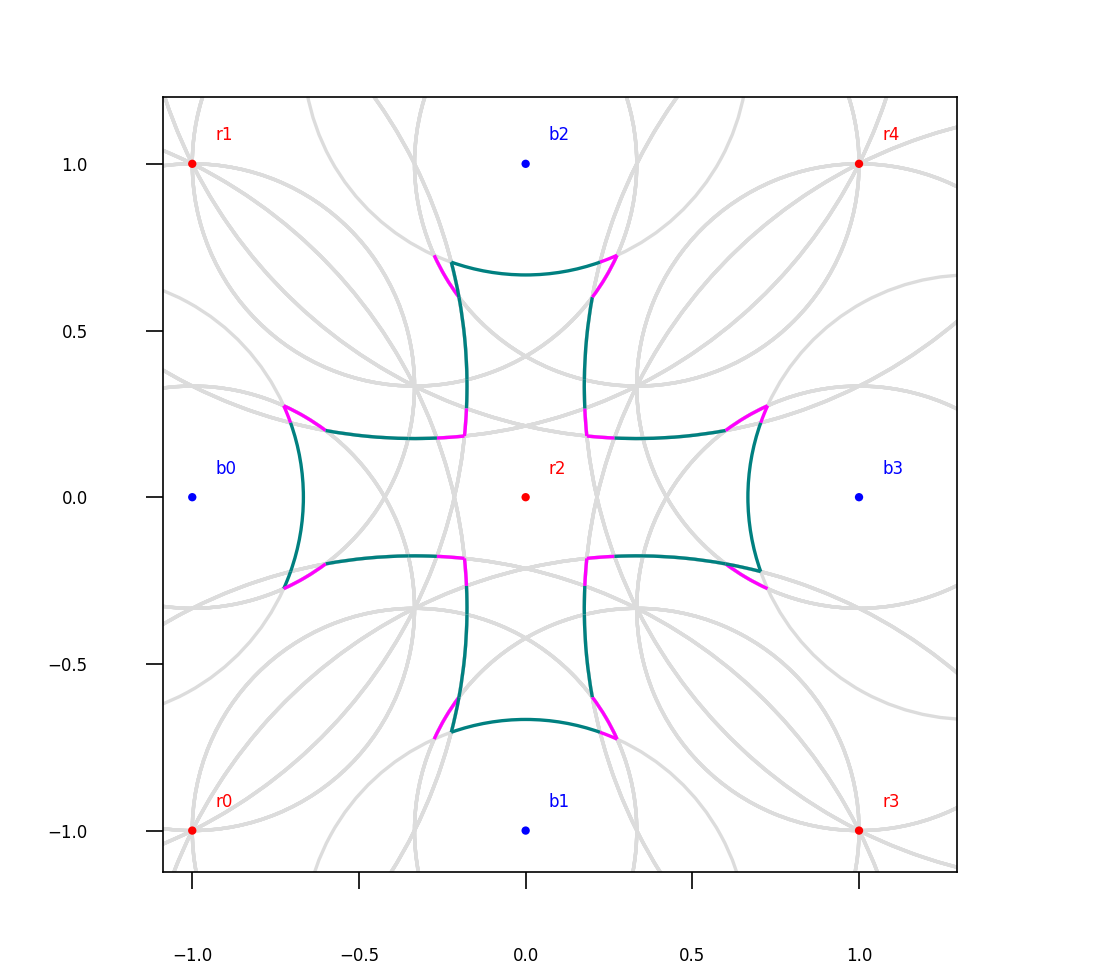}}\\
\subcaptionbox{\label{g3}}{\includegraphics[width = 0.5\columnwidth]{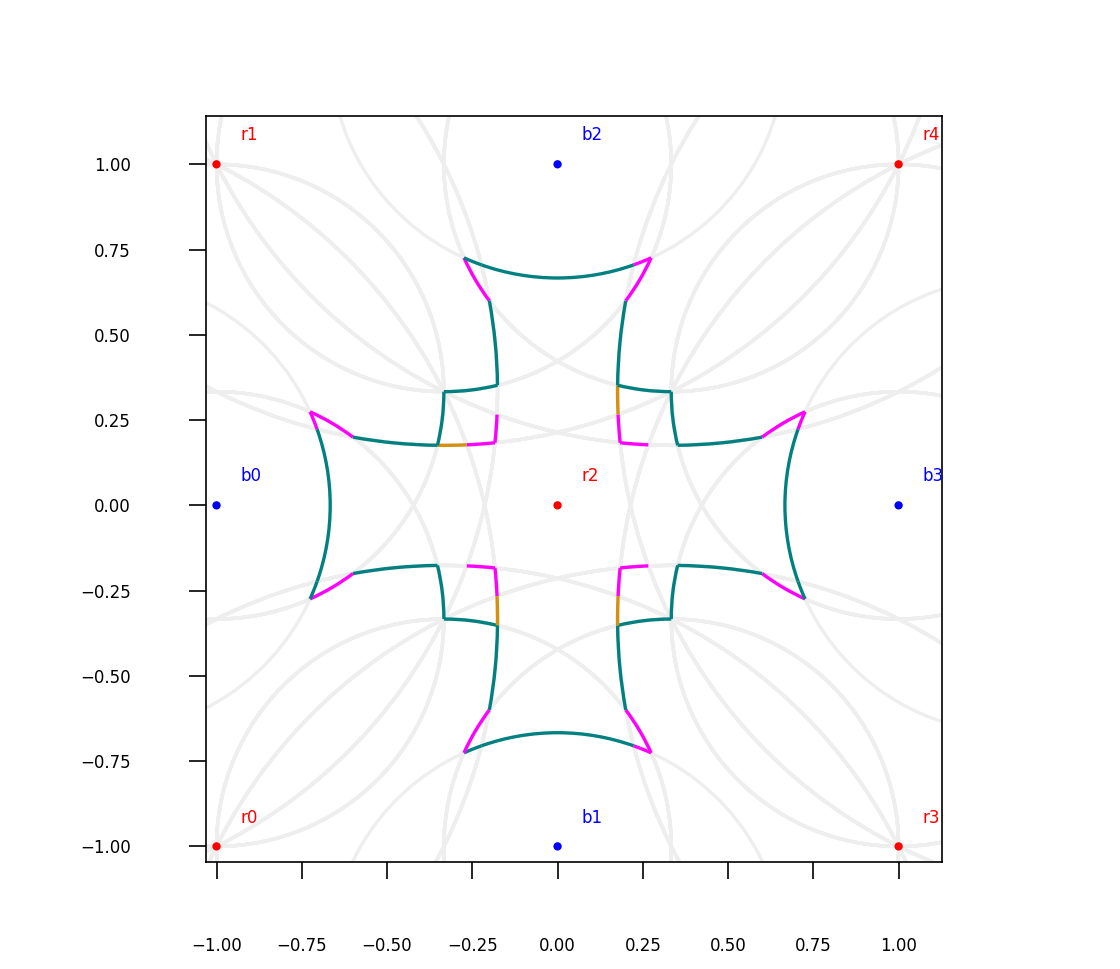}} &
\subcaptionbox{\label{g4}}{\includegraphics[width = 0.5\columnwidth]{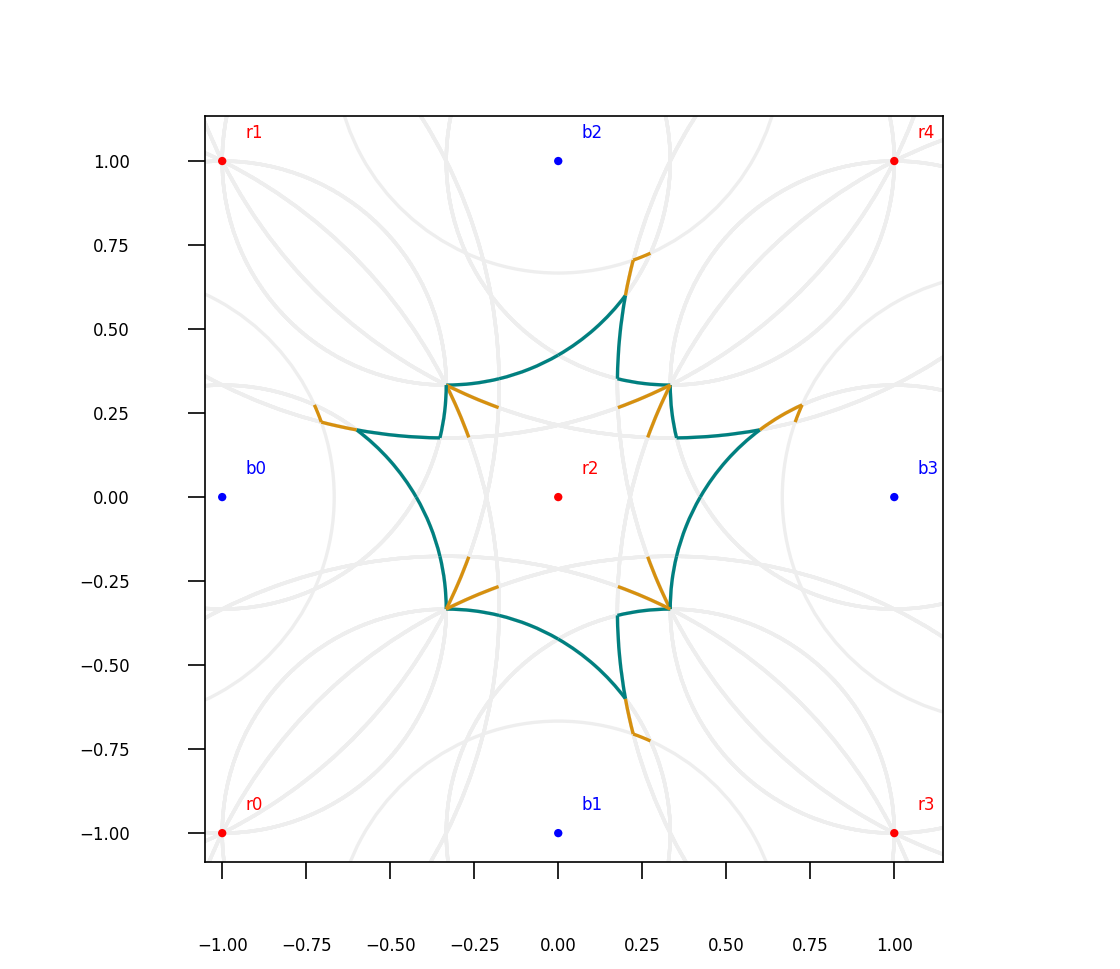}}
\end{tabular}
\caption{a) Points organised in a grid, with alternating red and blue points, and the resulting scenic graph. b) Route generated by the ACU algorithm on the graph in a). c) Route generated by the ACCH algorithm on the graph in a). d) Route generated by the DPE algorithm on the graph in a).}
\label{Grid_}
\end{figure}

\begin{table}[H]
\begin{center}
\caption{Results for experiments on graph in Fig. \ref{Grid_} (Grid) (Path length 558.96, No. of edges: 304)}\label{tab4}
\begin{tabular} {lllll}
\toprule
Algorithm & RL & NoE & NoRE & RE\%\\
\midrule
  ACU & 7.42 & 48 & 4 & 8.33\\
  ACCH & 7.7 & 48 & 12 & 25\\ 
  DPE & 6.15 & 36 & 16 & 44.44\\ 
 \bottomrule
 \end{tabular}
\end{center}
\end{table}

\begin{figure}
\begin{center}
\begin{tabular}{c}
\subcaptionbox{\label{l1}}{\includegraphics[width = 0.45\columnwidth]{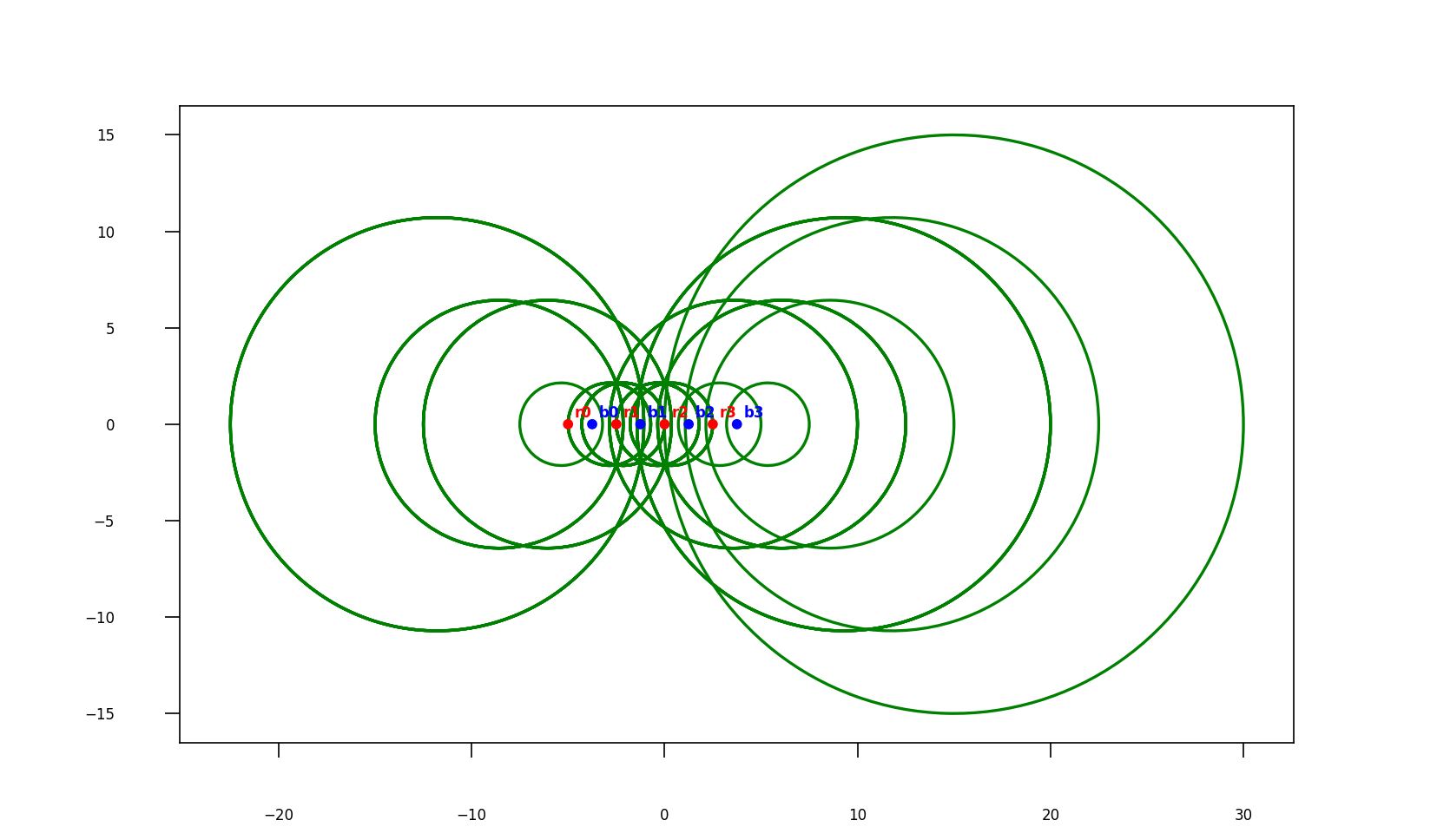}}\\
\subcaptionbox{\label{l2}}{\includegraphics[width = 0.45\columnwidth]{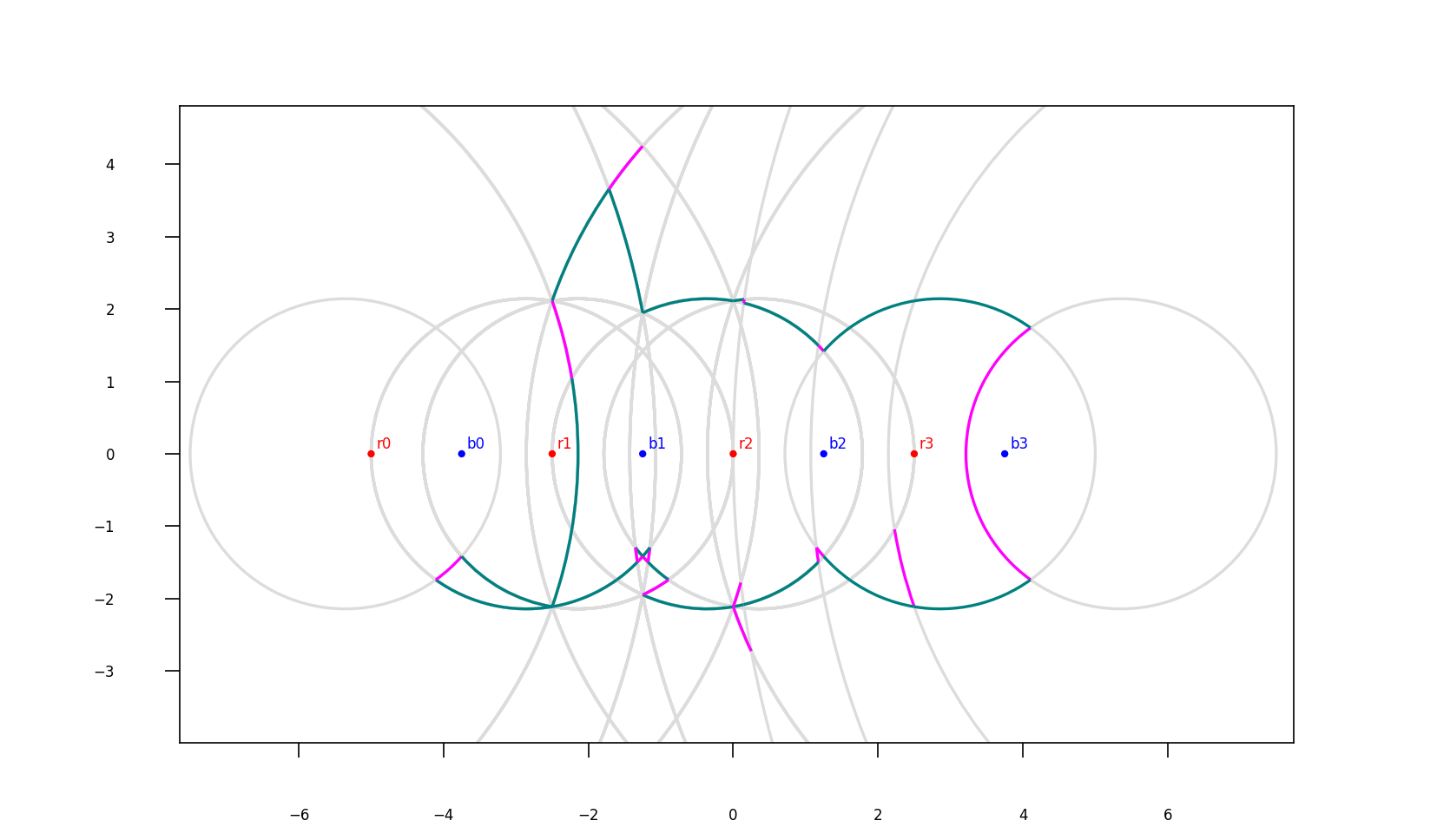}}\\
\subcaptionbox{\label{l3}}{\includegraphics[width = 0.45\columnwidth]{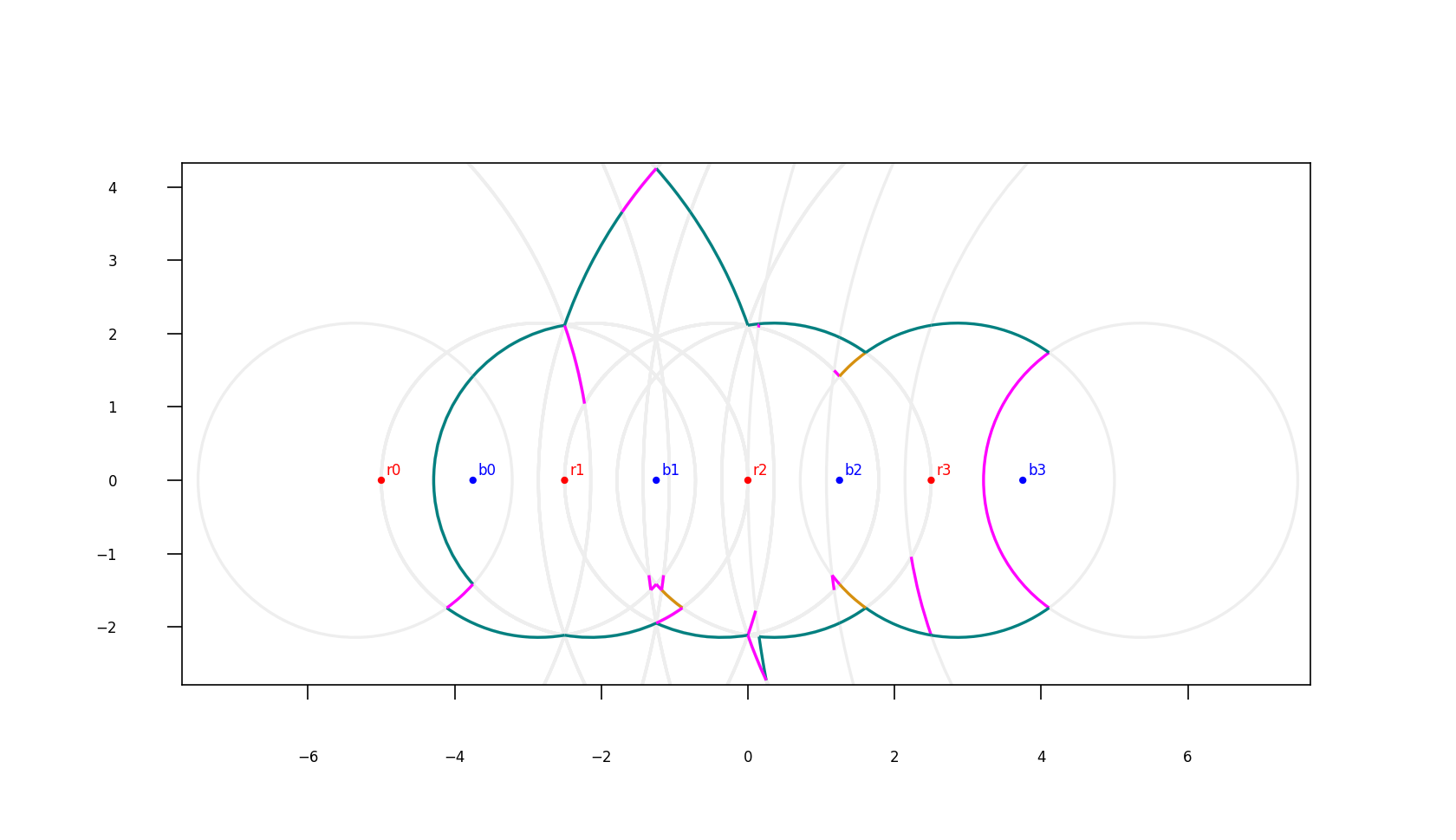}}\\
\subcaptionbox{\label{l4}}{\includegraphics[width = 0.45\columnwidth]{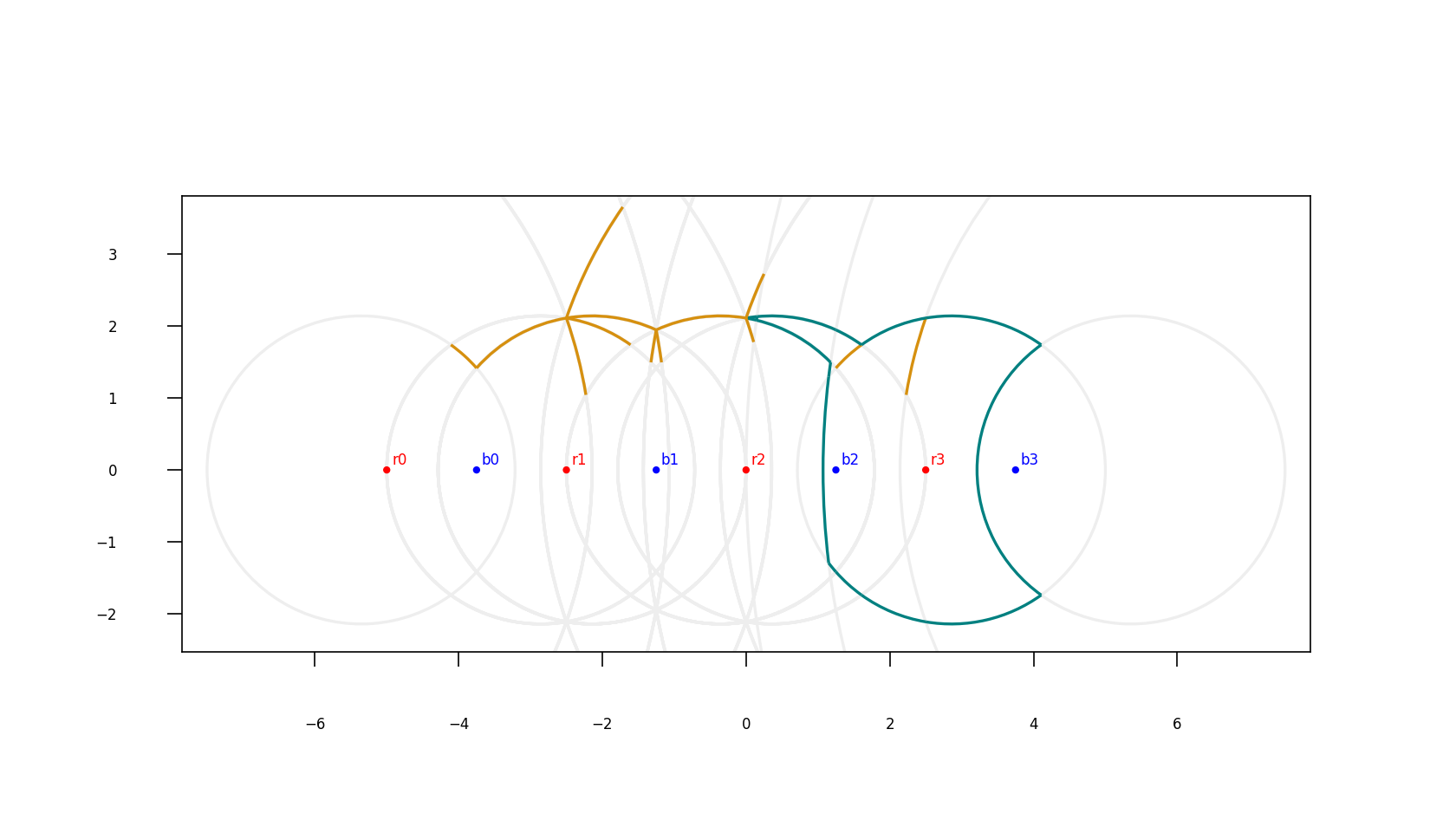}}
\end{tabular}
\caption{a) Points organised in a line, with alternating red and blue points. ($\delta = 0.75$), and the resulting scenic graph. b) Route generated by the ACU algorithm on the graph in a). c) Route generated by the ACCH algorithm on the graph in a). d) Route generated by the DPE algorithm on the graph in a).}
\label{line}
\end{center}
\end{figure}

\begin{table}[H]
\begin{center}
\caption{Results for experiments on graph in Fig. \ref{line} (Alternating red and blue points on line) (Path length 2100.38, No. of edges: 222)}\label{tab6}
\begin{tabular} {lllll}
\toprule
Algorithm & RL & NoE & NoRE & RE\%\\
\midrule 
  ACU & 33.55 & 39 & 4 & 10.25 \\
  ACCH & 33.63 & 37 & 15 & 40.54\\ 
  DPE & 27.77 & 28 & 13 & 46.42\\
 \bottomrule
 \end{tabular}
\end{center}
\end{table}

\begin{figure}
\begin{tabular}{cc}
\subcaptionbox{\label{c1}}{\includegraphics[width = 0.5\columnwidth]{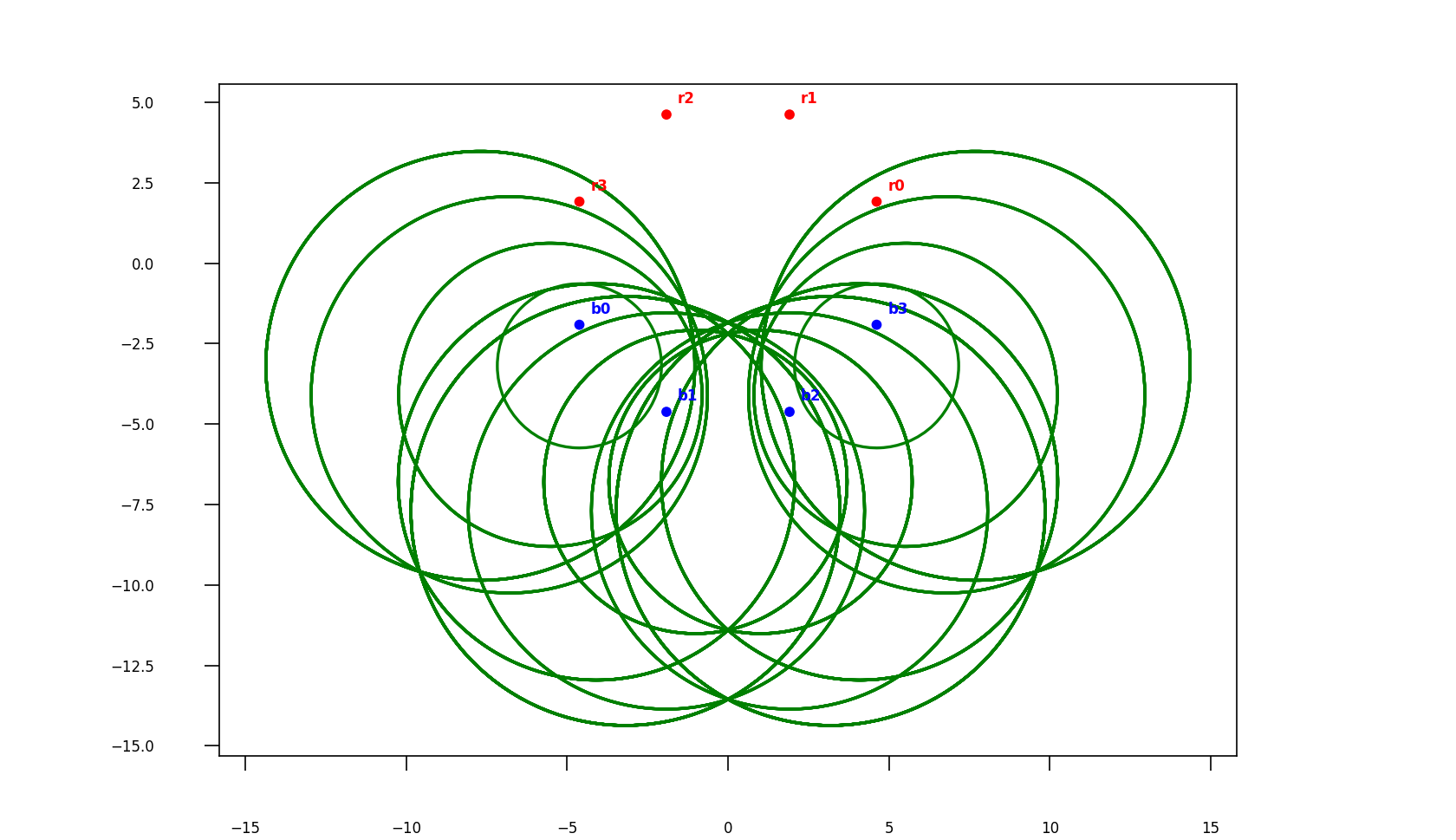}} &
\subcaptionbox{\label{c2}}{\includegraphics[width = 0.5\columnwidth]{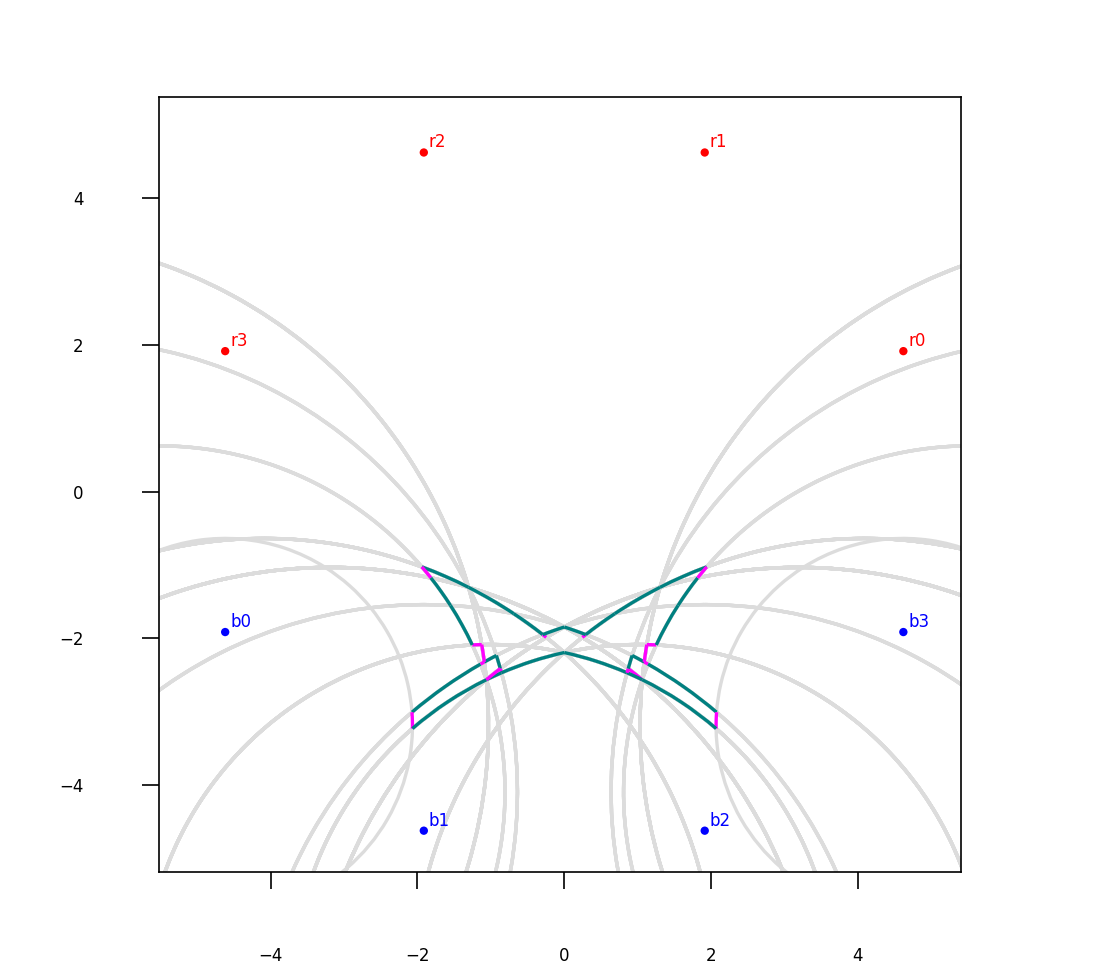}}\\
\subcaptionbox{\label{c3}}{\includegraphics[width = 0.5\columnwidth]{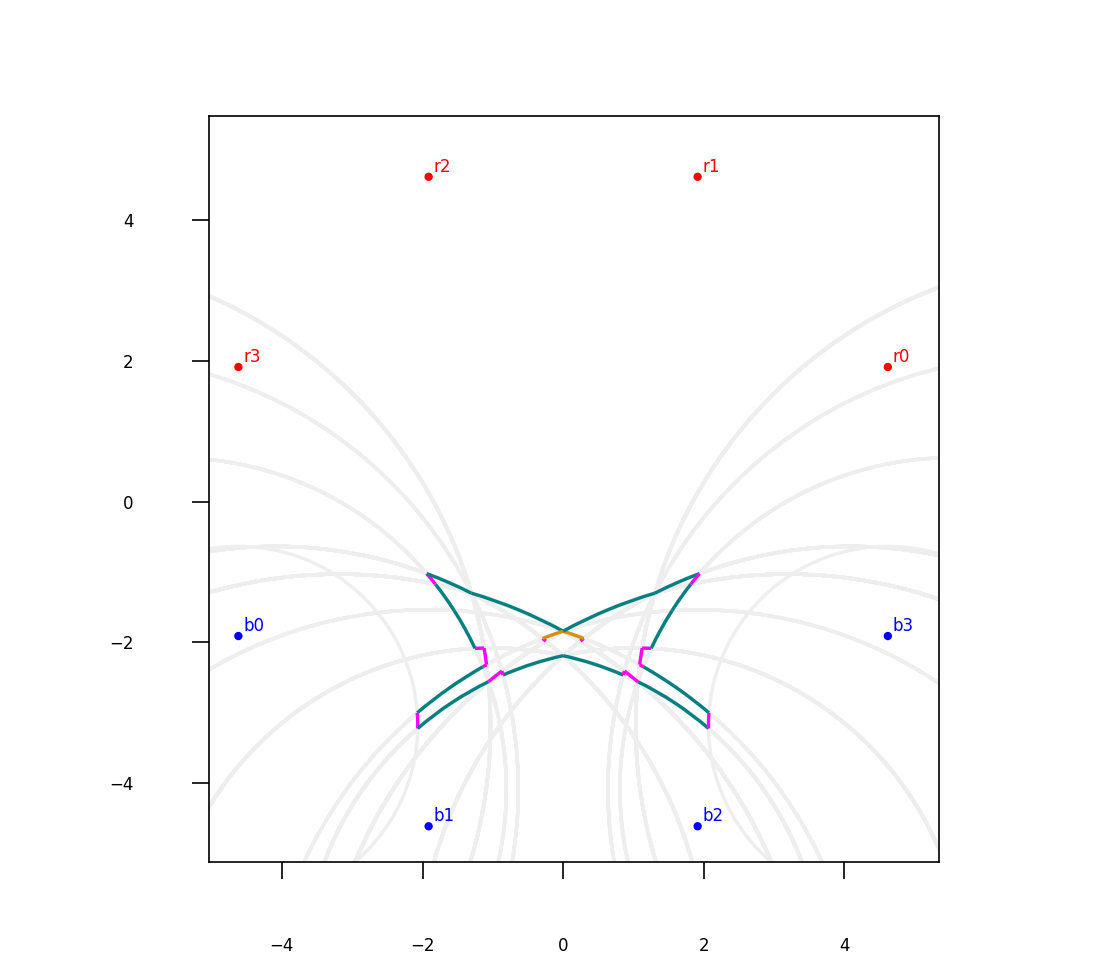}} &
\subcaptionbox{\label{c4}}{\includegraphics[width = 0.5\columnwidth]{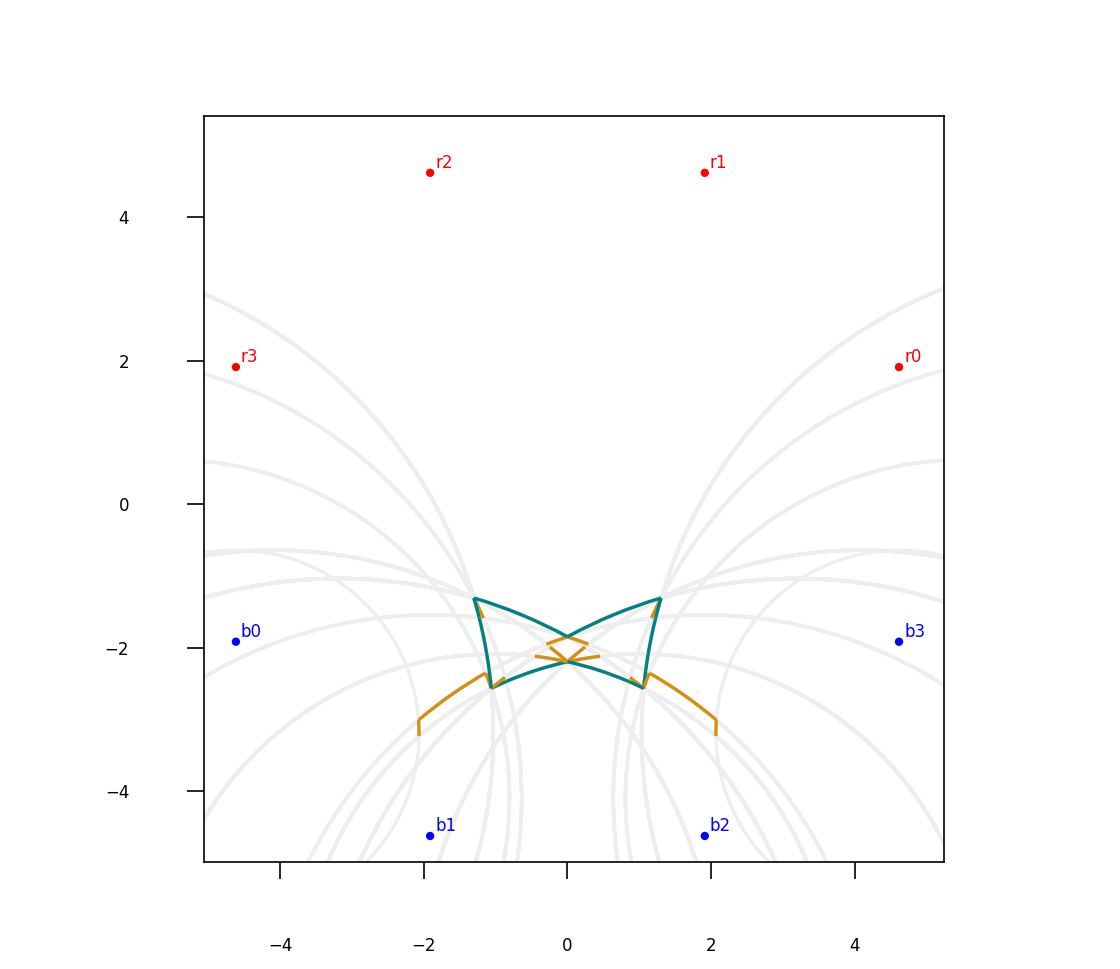}}
\end{tabular}
\caption{a) Points organised in a circle, with red and blue points forming a semicircle each, and the corresponding scenic graph. ($\delta = 0.5$), and the resulting scenic graph. b) Route generated by the ACU algorithm on the graph in a). c) Route generated by the ACCH algorithm on the graph in a). d) Route generated by the DPE algorithm on the graph in a).}
\label{circle_H}
\end{figure}

\begin{table}
\begin{center}
\caption{Results for experiments on graph in Fig. \ref{circle_H} (Half-circle red, Half-circle blue) (Path length 4175.04, No. of edges: 368)}\label{tab3}
\begin{tabular} {lllll}
\toprule
Algorithm & RL & NoE & NoRE & RE\%\\
\midrule
  ACU & 17.65 & 44 & 0 & 0\\
  ACCH & 15.91 & 32 & 4 & 12.5\\ 
  DPE & 14.13 & 30 & 16 & 53.33\\ 
 \bottomrule
 \end{tabular}
\end{center}
\end{table}

\begin{figure}
\begin{tabular}{cc}
\subcaptionbox{\label{h1}}{\includegraphics[width = 0.5\columnwidth]{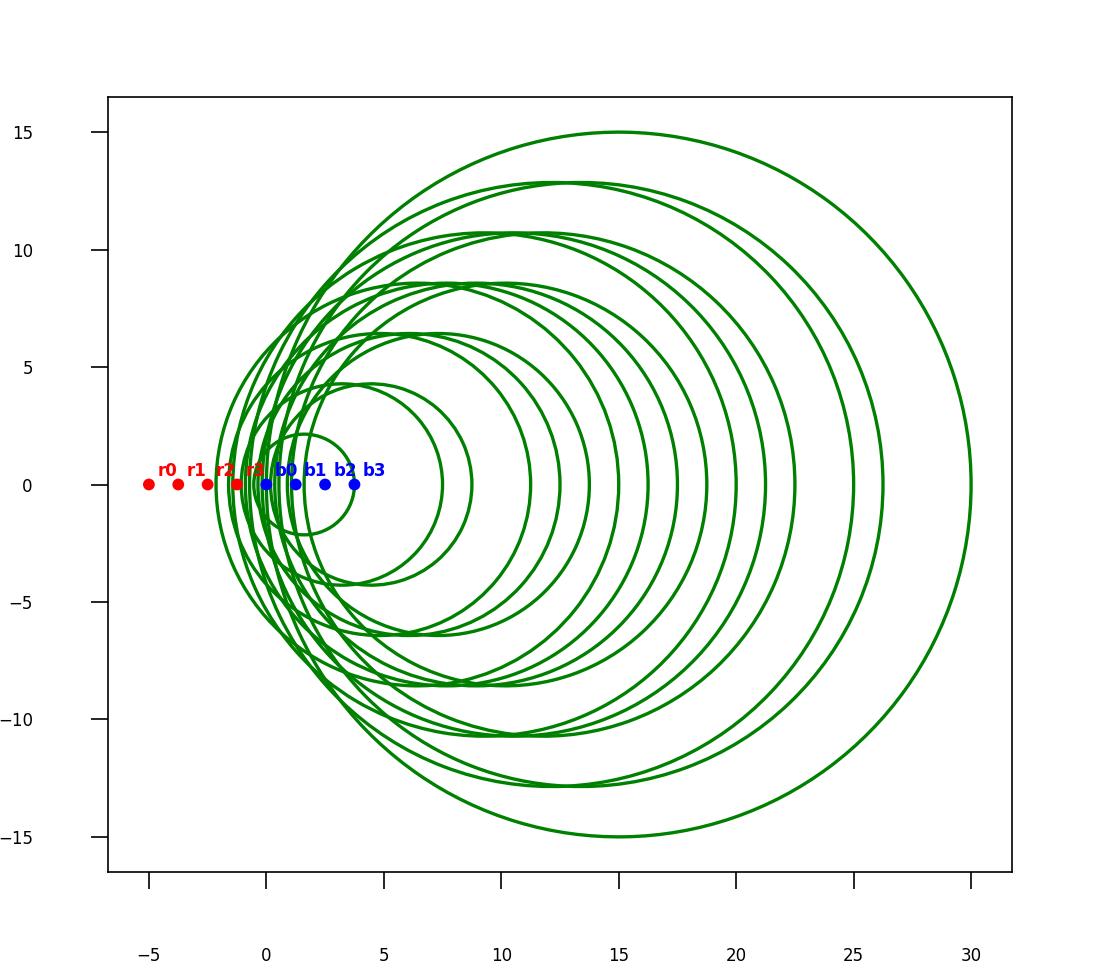}} &
\subcaptionbox{\label{h2}}{\includegraphics[width = 0.5\columnwidth]{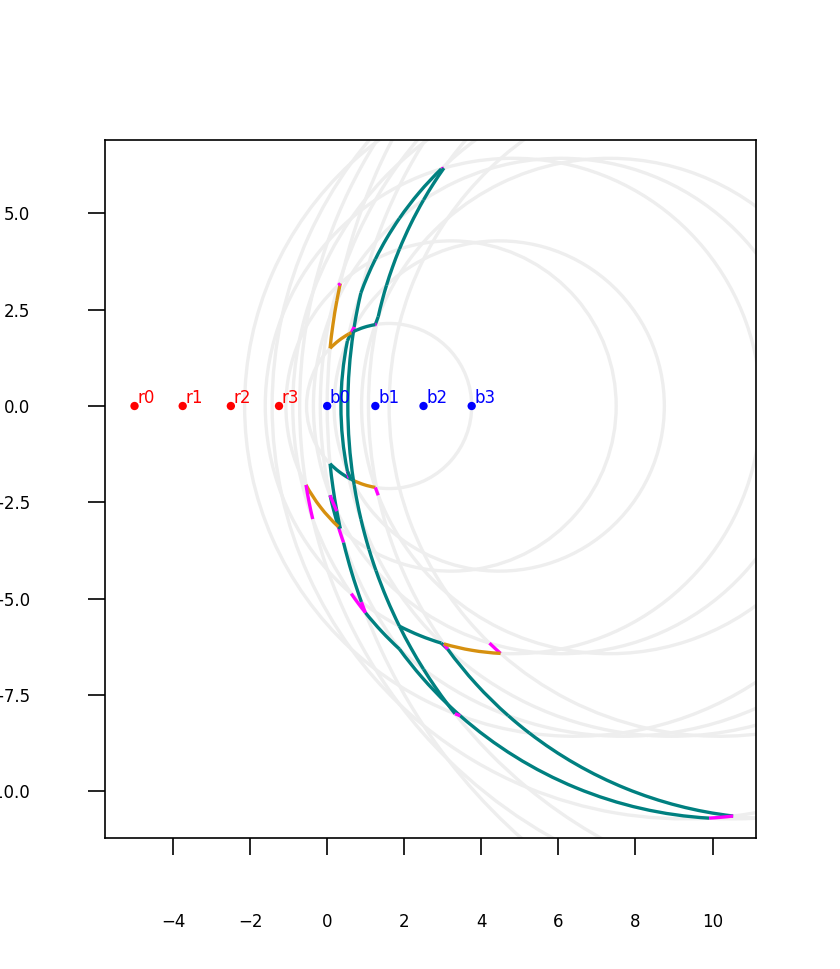}}\\
\subcaptionbox{\label{h3}}{\includegraphics[width = 0.5\columnwidth]{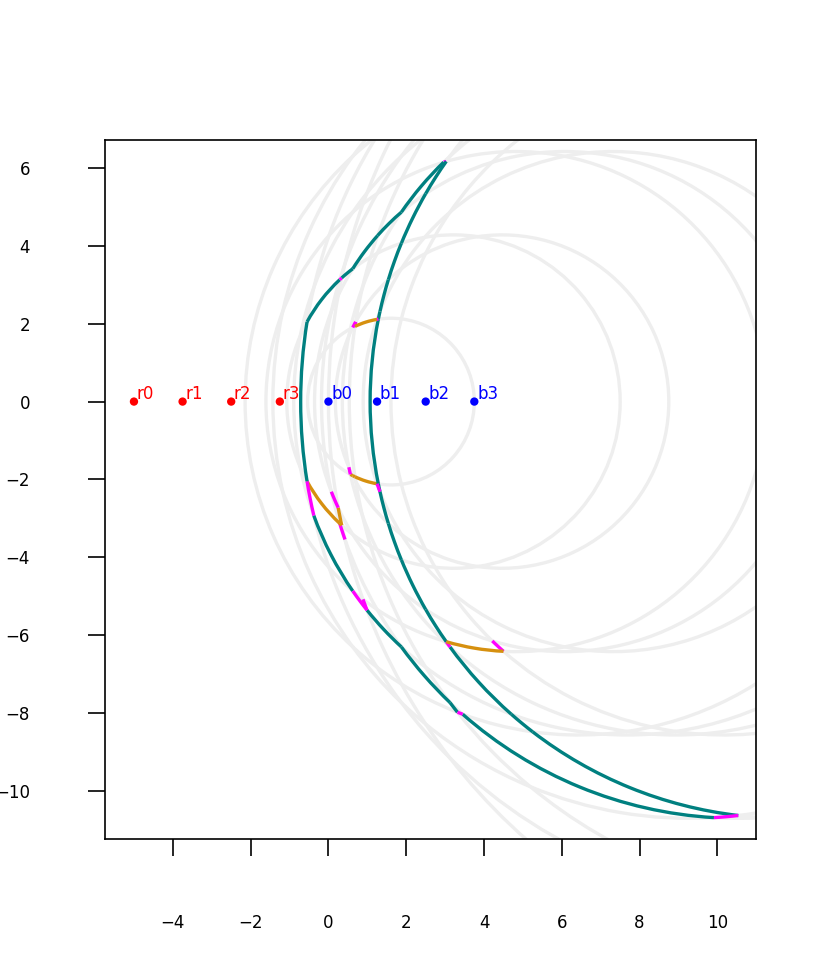}} &
\subcaptionbox{\label{h4}}{\includegraphics[width = 0.5\columnwidth]{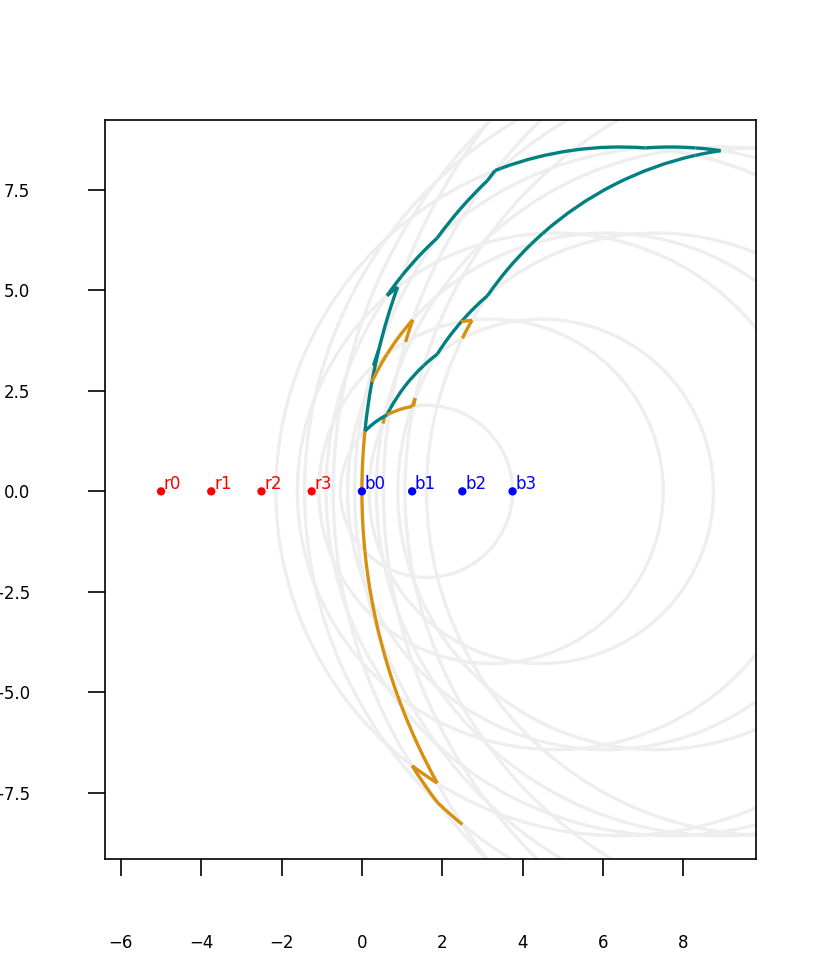}}
\end{tabular}
\caption{a) Points organised in a line, with half of the line consisting of red points and the other half consisting of blue points. ($\delta = 0.75$), and the resulting scenic graph. b) Route generated by the ACU algorithm on the graph in a). c) Route generated by the ACCH algorithm on the graph in a). d) Route generated by the DPE algorithm on the graph in a).}
\label{hline}
\end{figure}
\begin{table}
\begin{center}
\caption{Results for experiments on graph in Fig. \ref{hline} (Half-line red, Half-line blue) (Path length 861.69, No. of edges: 296)}\label{tab5}
\begin{tabular} {lllll}
\toprule
Algorithm & RL & NoE & NoRE & RE\%\\
\midrule
  ACU & 60.96 & 62 & 14 & 22.58 \\
  ACCH & 53.88 & 49 & 17 & 34.69\\ 
  DPE & 42.23 & 47 & 20 & 42.55\\ 
 \bottomrule
 \end{tabular}
\end{center}
\end{table}

\clearpage
\section{Some examples}
\label{Examples}
We now generate scenic routes on real-world examples by picking famous locations. In here we discard the notion of red-blue points, and we consider scenic points to satisfy the aforementioned condition for any two \textbf{points of interest}, that we mark in red. The ACU algorithm was used to generate routes for these scenic graphs.

\subsubsection*{The Pyramids of Giza}
\begin{figure}[h]
    \centering
    \begin{center}
      \includegraphics[width=0.7\textwidth]{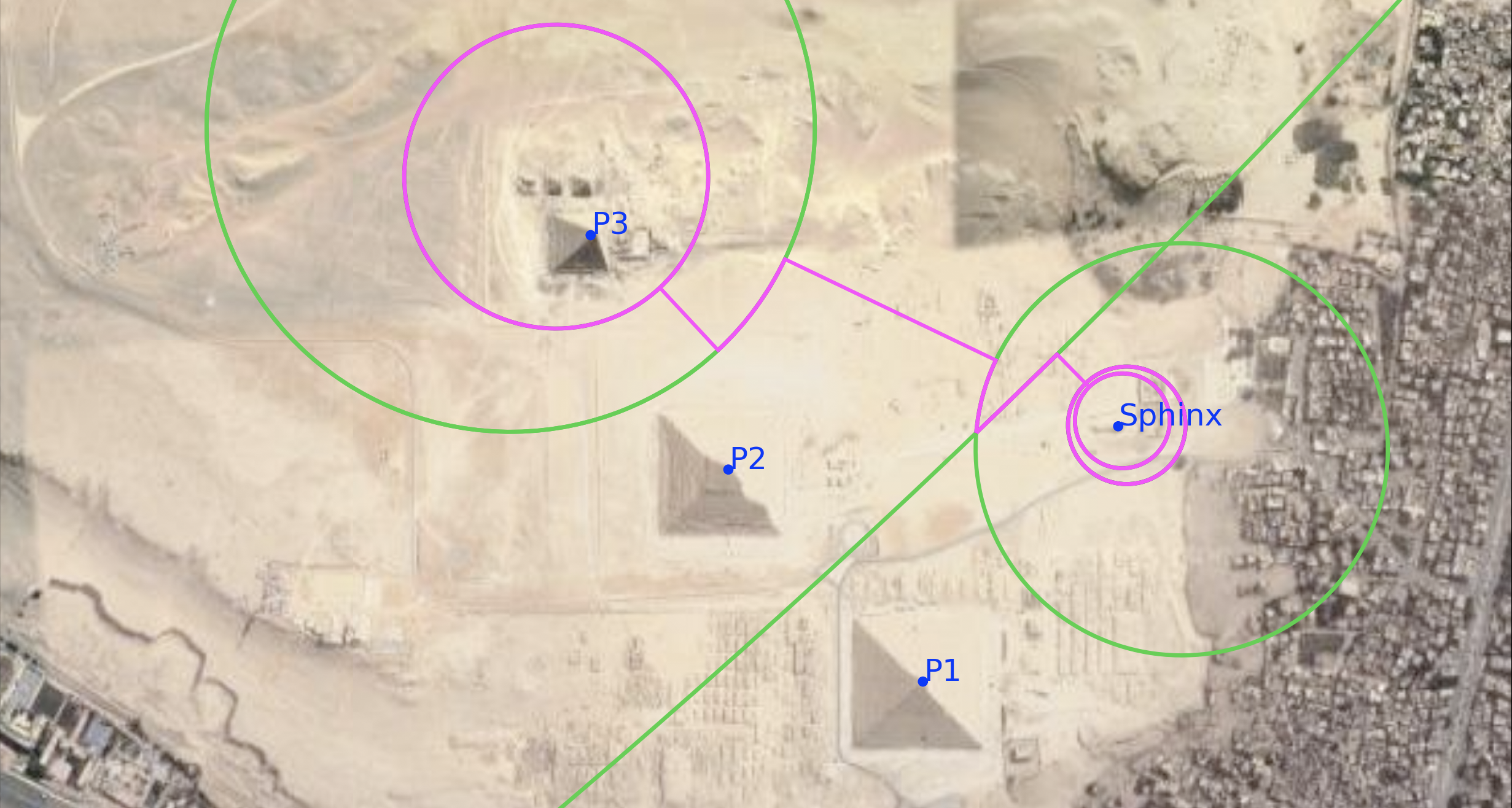}
    \end{center}
    \caption{A scenic route on the area of the Great Pyramids, generated using the ACU algorithm (The edges included in the generated route are in pink, and the rest of the scenic paths are in green.)}
    \label{fig-giza}
    \end{figure}

In Figure \ref{fig-giza}, the four point of interest that have been chosen are The Pyramid of Khufu (P1), The Pyramid of Khafre (P2), The Pyramid of Menkaure (P3), and The Great Sphinx. The weights of these points are set to their actual heights (P1: 138.5m, P2: 136.4m, P3: 61m, Sphinx: 20m).

\subsubsection*{The Eiffel Tower and surrounding areas}
\begin{figure}[h]
    \centering
    \begin{center}
      \includegraphics[width=0.5\textwidth]{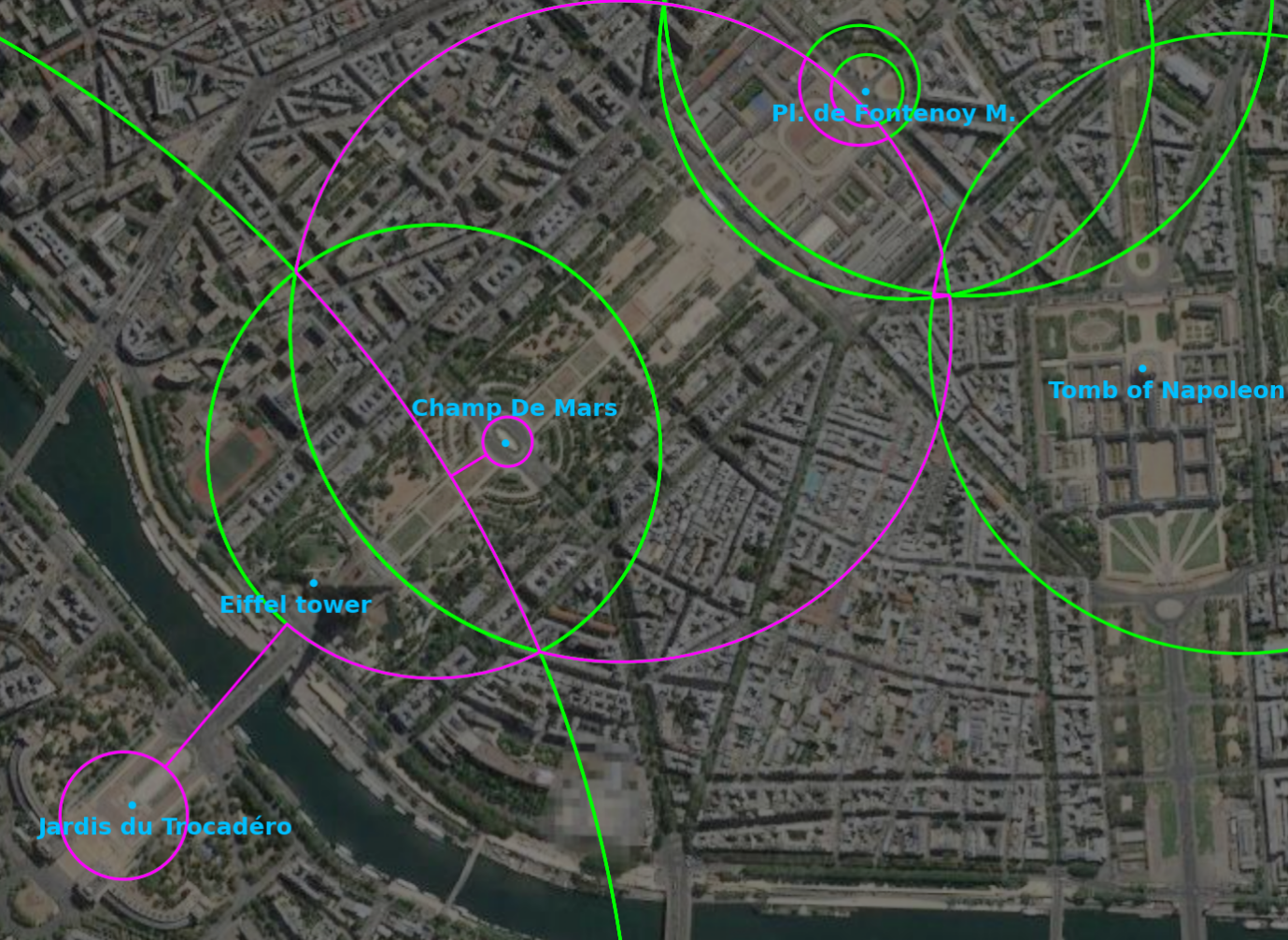}
    \end{center}
    \caption{A scenic route on the area of the Eiffel Tower, generated using the ACU algorithm (The edges included in the generated route are in pink, and the rest of the scenic paths are in green.)}
    \label{fig-paris}
    \end{figure}

In Figure \ref{fig-paris}, the four point of interest that have been chosen are The Eiffel Tower, The Jardis du Trocadéro, The Champ De Mars, The Tomb of Napoleon Bonaparte and the monument at the Place de Fontenoy square. Just like the previous example, the weights of these points are set to their actual heights (Eiffel Tower: 33m, Buildings at Jardis du Trocadéro: 70m, Champ De Mars: 20m, The Tomb of Napoleon Bonaparte: 107m the monument at Place de Fontenoy: 16m).

\section{Conclusion}
Given a set of red and blue points in 2D space, a scenic point is one whose distances from a particular red point and a particular blue point are in the ratio of the weight of the red point to that of the blue point. We introduce the concepts of scenic points, scenic paths and scenic routes in two-class weighted point configurations in 2D spaces, a characterization of the properties of a scenic route, and an analogy of these conditions in a real world scenario. We present three algorithms to generate scenic routes, and analyze the routes generated by these algorithms. We finally pick some real-world scenarios for which we generate scenic routes.

Our paper presents a preliminary idea of what a scenic point, path and route are. Further work may involve placing scenic points in a 3D space, new alternate definitions of what may constitute a scenic point, additional constraints on scenic routes and alternate algorithms that may focus on a different set of requirements.
\bibliographystyle{unsrt}
\bibliography{refs}
\nocite{*}

\end{document}